\documentclass[superscriptaddress,groupedaddress,nofootnoteinbib,12pt]{article}
\pdfoutput=1
\usepackage{color}
\usepackage{graphicx}
\usepackage{dcolumn}
\usepackage{bm}
\usepackage{amssymb}
\usepackage{amsmath}
\usepackage{sectsty}
\usepackage{colortbl}
\usepackage{latexsym}
\usepackage{float}
\usepackage{ifthen}
\usepackage{enumerate}
\usepackage{url}
\usepackage{jcappub}
    \usepackage{picinpar}
    \usepackage{colortbl}
\usepackage{multirow}
    \usepackage{float}
          \usepackage{setspace}
\usepackage{array}
\usepackage{bm}
\usepackage{amsopn}

\usepackage{booktabs}
\usepackage[table]{xcolor}
\definecolor{lightgray}{gray}{0.9}

\newcommand{\p}{\partial}

\def\ba{\begin{eqnarray}}
\def\ea{\end{eqnarray}}
\def\beq{\begin{eqnarray}}
\def\eeq{\end{eqnarray}}

\def\({\left(}
\def\){\right)}

\def\p{\partial}

\def\<{\langle}
\def\>{\rangle}

\usepackage{titlesec}
\titleformat{\part}{\Large\bfseries}{}{0pt}{Part \thepart\ --\ }

\DeclareRobustCommand{\rchi}{{\mathpalette\irchi\relax}}
\newcommand{\irchi}[2]{\raisebox{\depth}{$#1\chi$}}

\newcommand{\para}[1]{\par\vspace{2mm}\noindent\textbf{{#1}}.---}

\newcolumntype{Q}{>{$\displaystyle}l<{$}}
\newcolumntype{q}{>{\columncolor[gray]{0.9}$\displaystyle}l<{$}}
\newcolumntype{R}{>{$\displaystyle}r<{$}}
\newcolumntype{S}{>{$\displaystyle}c<{$}}
\newcolumntype{s}{>{\columncolor[gray]{0.9}$\displaystyle}c<{$}}
\newcolumntype{T}{>{\columncolor[gray]{0.9}}c<{}}

\newsavebox{\tableA}
\newsavebox{\tableB}

\newsavebox{\boxplot}
\newsavebox{\boxplota}

\definecolor{dullpurple}{rgb}{0.431,0.188,0.534}
\definecolor{darkgreen}{rgb}{0.133,0.545,0.133}

\begin{document}

\title{Quantum quenches during inflation}
   
\author{Pedro Carrilho$^{a}$ and Raquel H. Ribeiro$^{a}$}
\affiliation{$^{a}$Astronomy Unit, School of Physics and Astronomy, \\
Queen Mary University of London, \\
Mile End Road, London, E1 4NS, UK}

    \emailAdd{p.gregoriocarrilho@qmul.ac.uk}

\abstract{
We propose a new technique to study fast transitions during inflation, by studying the dynamics of quantum quenches in an $O(N)$ scalar field theory in de Sitter spacetime. We compute the time evolution of the system using a non-perturbative large-$N$ limit approach. We derive the self-consistent mass equation for several physically relevant transitions of the parameters of the theory, in a slow motion approximation. Our computations reveal that the effective mass after the quench evolves in the direction of recovering its value before the quench, but stopping at a different asymptotic value, in which the mass squared is strictly positive. Furthermore, we tentatively find situations in which the effective mass squared can be temporarily negative, thus breaking the $O(N)$ symmetry of the system for a certain time, only to then come back to a positive value, restoring the symmetry. We argue the relevance of our new method in a cosmological scenario.
}   

\maketitle

\section{Introduction}
\label{sec:introduction}

Inflationary cosmology has become the paradigm for understanding the early universe. This quasi-de Sitter period not only provides an explanation for the flatness and horizon problems, but also induces quantum fluctuations of one or more fields to grow and later seed the formation of structure. This proposal is strongly supported by observations of the Cosmic Microwave Background (CMB) most recently reported by the Planck collaboration~\cite{Ade:2015xua}, which are already able to accurately measure a number of parameters, such as the scalar spectral index $n_s$, and the amplitude of scalar perturbations~\cite{Ade:2015lrj}. These observations favor the simplest scenarios of early universe physics, in which inflation is modeled by a single, slowly-rolling scalar field with nearly gaussian statistics. Therefore, a number of more exotic scenarios have already been ruled out by these measurements, which demonstrates their ability to probe high energy physics in the early universe. Future experiments are expected to do even better, with hopes of pinning down the small primordial non-gaussianity at complementary scales through observations of the large scale structure~\cite{Amendola:2012ys,Maartens:2015mra,Abell:2009aa}.

On the other hand, another potential source of information at lower-order statistics is the presence of features in the primordial power spectrum of the gauge invariant curvature perturbation, $\zeta$~\cite{Bardeen:1980kt}. While most simple models predict a conserved $\zeta$ on super-horizon scales~\cite{Lyth:2004gb}, as well as a nearly scale-invariant power spectrum, $\mathcal{P}_\zeta\sim k^{n_s-1}$~\cite{Bardeen:1983qw}, a number of alternatives exist which can give rise to oscillations and other effects in the primordial signal. Such features can arise during short violations of slow-roll or more generally due to transient phenomena occurring during inflation. This has motivated work on extensions of the slow-roll formalism to more adequately deal with these events~\cite{Stewart:2001cd,Choe:2004zg,Dvorkin:2009ne}, as well as the development of numerical techniques to allow for multiple-field dynamics~\cite{Elliston:2013afa,Dias:2016rjq,Mulryne:2016mzv}. These phenomena can originate from various sources, such as fast changes in the speed of sound of perturbations~\cite{Chung:1999ve,Khoury:2008wj,Achucarro:2012fd,Achucarro:2013cva,Konieczka:2014zja,Mooij:2015cxa} as well as steps in the inflationary potential and its derivatives~\cite{Adams:2001vc,Joy:2007na,Ashoorioon:2006wc,Ashoorioon:2008qr,Hazra:2010ve,Adshead:2011jq}. These features can also arise directly from coherent oscillations of massive scalars~\cite{Chen:2014joa,Chen:2014cwa,Chen:2015lza,Chen:2016qce}. Looking for signatures of such violent dynamics can therefore hint at specific micro-physics processes which have occurred in the early universe and unveil more details about the physics of inflation. The experimental sensitivity to these features is expected to improve in large-scale-structure surveys in the near future~\cite{Chen:2016vvw}.

In this work we study fast phenomena which arise when there is an almost instant change of the couplings of the system during an inflationary phase---a \textit{quantum quench}. These phenomena are common in multiple field models where the inflationary trajectory suddenly turns in field configuration space as a result of the interaction between quantum fluctuations of different particle species. Generally, these fast events occur whenever there are very pronounced slopes in the potential which are traversed during very short times, $\Delta t\ll H^{-1}$. The end result is effectively a transition in the parameters of the potential, such as the masses and couplings of the fields. The interpretation of such features of the potential as quantum quenches is expected to be a good approximation for the description of the system some time after the violent phenomenon has occurred, while not depending on the exact details of the transition, provided that the transition is quicker than the other time scales of the system. This should allow for the study of the consequences of different classes of phenomena, based solely on the parameters of the potential before and after the transition has taken place. This generality is a major motivation for our approach.

Quantum quenches have been extensively studied in flat spacetime in the context of condensed matter physics (see for example Refs.~\cite{Calabrese:2006rx,Sotiriadis:2010si,Hung:2012zr} and references therein). Specifically, it was found that in flat space the system retains some memory of the conditions before the violent event, in the form of a dependence of the post-quench mass on the mass of the field before the quench~\cite{Hung:2012zr}. These techniques have also been used in a cosmological context to treat fast phase transitions in the early universe, in which the temperature is believed to decrease swiftly enough for the transition to the broken phase to be modeled by a quench in flat spacetime, such as the study of baryogenesis during the electroweak phase transition~\cite{Tranberg:2006dg} or the study of tachyonic preheating~\cite{Arrizabalaga:2004iw}. The quench approximation has also been used to study phase transitions in the context of inflation and its end~\cite{Boyanovsky:1996rw,Boyanovsky:1996sq,Boyanovsky:1996fz,Boyanovsky:1996sv,Boyanovsky:1997cr,Boyanovsky:1997xt,Boyanovsky:2006bf}, in which case a de Sitter background was used. While unrelated to cosmology, studies targeting the  universality of fast quenches have also been made in curved spacetime backgrounds~\cite{Das:2014hqa}. In spite of all these applications, this technique is yet to be applied to the study of general transitions which do not change the phase of the system. This is the main aim of our work. We also focus on the comparison with the flat situation, and in particular in understanding the role of the curved spacetime in the structure of a theory after it has undergone a quantum quench.

More specifically, we study quenches of an $O(N)$ scalar field theory on a de Sitter background in the so-called large-$N$ approximation~\cite{Cooper:1987sa,Cooper:1994hr,Cao:2001hn,Moshe:2003xn}. This method allows one to study a theory with a large number, $N$, of identical fields by expanding the action in powers of $1/N$, instead of the usual expansion in powers of the coupling constant. Consequently, this is a manifestly non-perturbative method since it allows for studying systems with large couplings. Already in flat spacetime such a problem is hard to resolve mainly because perturbation theory breaks down at the onset of the system becoming strongly coupled. This means that whichever observable quantities one wishes to compute using Feynman diagrams, one would have to solve an infinite number of these diagrams corresponding to all orders in perturbation theory to obtain a meaningful result.

In the context of condensed matter physics, the large-$N$ treatment has been quite successful, as it has been shown to greatly simplify the calculation of the effects of interactions, essentially reducing to finding a solution of a self-consistent equation for the effective mass of the fields. Furthermore, the large-$N$ expansion and other non-perturbative techniques are very useful in describing IR effects in de Sitter, as is evidenced by Refs.~\cite{Riotto:2008mv,Serreau:2011fu,Parentani:2012tx,Starobinsky:1994bd,Cooper:1986wv,Cao:2004mn,Gautier:2015pca}. Ref.~\cite{Serreau:2011fu}, for example, uses large-$N$ to show that IR effects and self-interactions force the effective mass of the fields to be strictly positive, something that had already been discussed in the stochastic context~\cite{Starobinsky:1994bd}. This effect, which is proportional to the root of the coupling constant, $\sqrt{g_4}$, would be impossible to obtain using perturbative methods. The subject of IR effects and dynamical mass generation has also been studied in Euclidean de Sitter spacetime~\cite{Rajaraman:2010xd,Beneke:2012kn} in which it has been extended to include higher orders in the large-$N$ expansion~\cite{Nacir:2016fzi}.

To summarize, in this work we apply the large-$N$ expansion to the quench of a scalar field theory, in a de Sitter background. We study the consequences of the quench for the evolution of the system, taking into account IR effects. We also compare the effects of introducing a quench in flat spacetime and de Sitter spacetimes.
Among the main objectives of this paper are the evaluation of the constraints found by previous authors on (the sign of) the effective mass in the unquenched situation and on the memory of the system in the flat case. For these applications it suffices to study the system in a static de Sitter background. Our future goal is, however, to calculate the observable consequences of the quench during slow-roll inflation, which will induce a change from the static situation to a dynamical one, and to calculate the leading order slow-roll correction to the power spectrum. This is phenomenologically motivated, since it can have measurable effects on observable quantities, such as the power spectrum and higher-order correlation functions (as has been argued in various contexts in cosmology, e.g.~Refs.~\cite{Lalak:2007vi, Burrage:2011hd, Avgoustidis:2011em, Ribeiro:2012ar, Dias:2012qy}).

The paper is organized as follows. In section \ref{sec:largeN}, we start by introducing our model and reviewing its large-$N$ expansion in de Sitter, in the unquenched case. We then include the quench and compute the two-point function after the quench in section \ref{sec:quenches}. In the first part of that section, we present analytical estimates for the evolution and late-time limit of the effective mass and then we move on to our numerical results to verify and correct our analytical calculations. We conclude in section \ref{sec:Discussion}, by enumerating the advantages of our approach.

\section{Large-$N$ in de Sitter}
\label{sec:largeN}

The action for an $N$-component $\phi^4$ model in a de-Sitter background geometry in $d$ spacetime dimensions is given by
\begin{equation}
S[\phi]=\int{\text{d}^dx\sqrt{-g}\left[-\frac12g^{\mu\nu}\p_\mu\phi^a\p_\nu\phi^a-\frac12\mu^2\phi^a\phi^a-\frac{g_4}{4N}(\phi^a\phi^a)^2 \right]}\,,
\end{equation}
where $g$ is the determinant of the metric, $a$ is an $O(N)$ index which labels the field (not to be confused with the scale factor) and repeated indices are summed over as per Einstein's notation. The metric is written in terms of conformal time $\tau$,
\begin{equation}
g_{\mu\nu}=a(\tau)^2\eta_{\mu\nu}\,,
\end{equation}
in which $\eta_{\mu\nu}$ is the Minkowski metric with mostly plus signature. For exact de-Sitter one has
\begin{equation}
a(\tau)=-\frac1{H\tau}\,,
\end{equation}
with $H$ being the Hubble parameter and $-\infty<\tau<0$ being the conformal time.

We now review the large-$N$ approximation. The general idea is that for a very large number of fields, $N\gg 1$, the action becomes very large, i.e. $S\gg\hbar$. As a consequence, the path integral is dominated by  solutions which minimize the action, just as it happens when one takes the classical limit ($\hbar\rightarrow 0$). This simplifies a number of calculations while still keeping contributions of all orders in the couplings of the theory. To see this explicitly, let us start by writing the path integral in the \textit{in-in} formalism as
\begin{equation}
\int_{\text{CTP}}\mathcal{D}\phi\ e^{iS[\phi]}\,,
\end{equation}
in which CTP is designating the closed-time-path measure one uses to account for the boundary conditions of the \textit{in-in} formalism. We now introduce a new variable defined by
\begin{equation}
\rho\equiv\phi^a\phi^a/N\,,
\end{equation}
which symbolizes the variance of the fields.
We can also change the path-integral by using the identity
\begin{equation}
\mathbf{1}\sim\int{\mathcal{D}\rho \ \delta(\phi^a\phi^a-N \rho)}\sim\int{\mathcal{D}\rho\,\mathcal{D}\xi\ e^{-\frac{i}2\int \text{d}^dx\sqrt{-g}\xi(\phi^a\phi^a-N \rho)}}\,,
\end{equation}
which results in
\begin{equation}
\int_{\text{CTP}}{\mathcal{D}\phi\mathcal{D}\rho\mathcal{D}\xi\ e^{iS[\phi,\rho,\xi]}}\,,
\end{equation}
where the new action $S[\phi,\rho,\xi]$ is given by
\begin{equation}
S[\phi,\rho,\xi]=\int{ \text{d}^dx\sqrt{-g}\left[-\frac12g^{\mu\nu}\p_\mu\phi^a\p_\nu\phi^a-\frac12(\mu^2+\xi)\phi^a\phi^a-\frac{Ng_4}{4}\rho^2+\frac{N}{2}\xi\rho \right]}\,.
\end{equation}
It is clear that the action above is simply quadratic in $\phi^a$, which allows one to perform $N$ Gaussian integrals for each field.
Before that, however, it is convenient to change variables to
\begin{equation}
\phi^a \equiv \rchi^a \, \displaystyle{a^{\frac{2-d}{2}}}\,,\ \ \ \ \ \rho \equiv \Pi \, a^{2-d}\,,
\end{equation}
since it is $\rchi^a$ which is the canonically normalized field in a de Sitter spacetime.
Integrating out $N-1$ copies of the $\rchi^a$ fields and substituting for the de Sitter metric, yields the following path integral
\begin{equation}
\int_{\text{CTP}}{\mathcal{D}\Pi\, \mathcal{D}\xi\, \mathcal{D}\sigma\ e^{iS_{\text{eff}}[\Pi,\xi,\sigma]}}\,,
\end{equation}
with
\begin{gather}
S_{\text{eff}}[\Pi,\xi,\sigma]=\int{ \text{d}^dx \left\{\frac12\sigma \left[\p^2+\frac{1}{\tau^2}\left(\frac{d(d-2)}{4}-\frac{\mu^2+\xi}{H^2}\right)\right]\sigma+N\left(\frac{\xi\Pi}{2(H\tau)^2} -\frac{g_4}{4}\Pi^2 (-H\tau)^{d-4}\right)\right\}}\nonumber\\
+(N-1)\frac{i}{2}\text{Tr}\log{\left[-\p^2-\frac{1}{\tau^2}\left(\frac{d(d-2)}{4}-\frac{\mu^2+\xi}{H^2}\right)\right]}\,,
\end{gather}
in which Tr is the functional trace defined by
\begin{equation}
\text{Tr}[f(x,y)]=\int{ \text{d}^dx \, f(x,x)}\,,
\end{equation}
and $\p^2$ is the Minkowski Laplacian. We have not integrated one of the scalar fields, given by $\sigma\equiv\rchi^N=\phi^N/(- H \tau)$, should there be a spontaneous breaking of the $O(N)$ symmetry, in which case $\sigma=O(\sqrt{N})$ instead of $O(1)$, as is assumed for all other field components, $\chi^a$. Should that be the case, it is clear that all terms in the action are order $N$ and thus, in the large-$N$ limit, one has $S\propto N \gg \hbar$. The path integral can then be evaluated by simply using the stationary phase approximation. Therefore, one must only minimize the action by imposing the following conditions with respect to each of the field species present:
\begin{gather}
\frac{\delta S_{\text{eff}}}{\delta\xi}=0\Rightarrow \frac{\bar\Pi}{(H\tau)^2}-\frac{\bar\sigma^2}{(H\tau)^2}+i\frac{\delta}{\delta\xi}\text{Tr}\log{\left.\left[-\p^2-\frac{1}{\tau^2}\left(\frac{d(d-2)}{4}-\frac{\mu^2+\xi}{H^2}\right)\right]\right|_{\xi=\bar\xi}}=0\,,\\
\frac{\delta S_{\text{eff}}}{\delta\Pi}=0\Rightarrow \frac{\bar\xi}{(H\tau)^2}-g_4\bar\Pi(-H\tau)^{d-4}=0\,,\\
\frac{\delta S_{\text{eff}}}{\delta\sigma}=0\Rightarrow\left[\p^2+\frac{1}{\tau^2}\left(\frac{d(d-2)}{4}-\frac{\mu^2+\bar\xi}{H^2}\right)\right]\bar\sigma=0\,.
\end{gather}
The barred variables ($\bar\Pi$, $\bar\xi$, $\bar\sigma$) denote the solutions to these equations of motion. For the case of $\bar\sigma$ we also factor out $\sqrt{N}$, for clarity of presentation.\footnote{Please note that should we be dealing with the $O(N)$ symmetric phase, we will simply set $\bar\sigma=0$, as it is assumed to be order $1/\sqrt{N}$ and hence negligible in the large-$N$ limit. For the broken phase, it is order $1$.} Using the first two equations above, gives rise to the following self-consistent equation for the effective mass
\begin{equation}
\label{selfconsdS}
m^2=\mu^2+\bar\xi=\mu^2+g_4(-H\tau)^{d-2}\left[\bar\sigma(x)^2+i G(x,x)\right]\,,
\end{equation}
where $G(x,x)$ is the Green's function of $\rchi^a$ evaluated at the same spacetime point, $x$, which can be calculated as an integral over the power spectrum:
\begin{equation}
G(x,x)=\int{\frac{\text{d}^{d-1}k}{(2\pi)^{d-1}}\ \tilde G (\tau,\tau,k)}\,.
\end{equation}
Note that the relation given by Eq.~\eqref{selfconsdS} depends non-trivially on the mass $m^2$ due to contributions from $G$ and $\bar\sigma$. Therefore, it allows one to find the effective mass which consistently includes all contributions from the interaction terms. The importance of the equal time propagator, $\tilde G(\tau,\tau,k)$ cannot be overstated: it is the main object that encodes the details of the interactions and therefore allows one to calculate the effective mass. As a result, the power spectrum will be the main object of focus with regards to the cosmological implications of our work. Much of the following sections is dedicated to its calculation.

\subsection{No quench}\label{NoQuench}

Before evaluating the consequences of a quench in this system, let us look at the simpler case in which there are no sudden changes in the parameters. This will serve to set some of the notation and also to explain the general procedure.

Our aim is to make use of Eq. \eqref{selfconsdS} to calculate the effective mass in the limit in which the mass is small, i.e. when $m/H\ll1$. This is the interesting case, since the effects of the curved background would disappear should one take the opposite limit.
The first step is the calculation of the Green's function. That can be easily done by expanding the fields in Fourier space, in terms of creation and annihilation operators, $a^\dagger_k$ and $a_k$,\footnote{Note that we are using an unlabeled field, $\rchi$, to represent each of the fields $\rchi^a$. We also omit the $O(N)$ indices everywhere else to avoid clutter.}$^{,}$\footnote{Note that, in general, the expansion of multiple interacting fields in creation and annihilation operators is not diagonal, i.e. each field depends on all of the $N$ pairs of ladder operators and not just on one of them, as seen here. The simplicity of the case presented here is due to the fact that the fields are effectivelly free in the large-$N$ limit, since all the effects of the interactions are contained in the effective mass. Thus it is possible to expand each field with just one pair of creation and annihilation operators, as shown in Eq.~\eqref{expaad}.}
\begin{equation}
\label{expaad}
\chi(\tau,\vec{x})=\int{\frac{\text{d}^{d-1}k}{(2 \pi)^{d-1}}\left(a_k u_k(\tau) e^{i\vec{k}\cdot\vec{x}}+a_k^\dagger (u_k(\tau))^* e^{-i\vec{k}\cdot\vec{x}}\right)}\,,
\end{equation}
in which $a^\dagger_k$ and $a_k$ obey the standard commutation relations:
\begin{equation}
[a_k,a_q^\dagger]=(2\pi)^3\delta^{(3)}(\vec k-\vec q)\,,\ [a_k,a_q]=0\,,\ [a_k^\dagger,a_q^\dagger]=0\,.
\end{equation}
The computation of the two-point function at the same point is straightforward, being given by
\begin{equation}
\label{twopointunquench}
\<0|\chi(\tau,\vec{x})\chi(\tau,\vec{x})|0\>=\int\frac{\text{d}^{d-1}k}{(2 \pi)^{d-1}}\left|u_k(\tau)\right|^2\,.
\end{equation}
This simply depends on the normalized wave-functions $u_k(\tau)$, which can be obtained from the Klein--Gordon equation, assuming the effective mass is constant.\footnote{This assumption is well motivated in a de Sitter invariant state, given that in that situation the two-point function for $\phi$ is constant \cite{Serreau:2011fu}, implying that $G(x,x)\propto (H\tau)^{-2}$. The field $\bar\sigma$ has the same behavior in such a state. This is purely a consequence of the de Sitter symmetry~\cite{Weinberg:2010fx}.} 
Choosing the Bunch--Davies vacuum, the wave-functions are given by
\begin{equation}
\label{wavfunc}
u_k(\tau)=-\frac12\sqrt{\frac{\pi}{2}}(1+i)e^{\frac{i\pi\nu}{2}}\sqrt{-\tau}\, H^{(1)}_\nu(-k\tau)\,,
\end{equation}
in which $H^{(1)}_\nu$ is the Hankel function of the first kind and $\nu$ is related to the mass of the field via
\begin{equation}
\nu=\sqrt{\left(\frac{d-1}{2}\right)^2-\frac{m^2}{H^2}}\,.
\end{equation}
The self-consistency condition, Eq. \eqref{selfconsdS}, then translates to, in $d=4$,
\begin{equation}
m^2=\mu^2+g_4(-H\tau)^{2}\left[\bar\sigma^2+\int{\frac{\text{d}^{3}k}{(2\pi)^{3}}\frac{\pi}{4}(-\tau)\left|H^{(1)}_\nu(-k\tau)\right|^2 }\right]\,.
\end{equation}
The integral on the r.h.s. is not straightforward to calculate analytically for a general order of the Hankel function. Furthermore, it has UV divergences which need to be regularized. These two issues are discussed, for example, by Serreau \cite{Serreau:2011fu}, and we shall follow the same procedures:
\begin{itemize}
\item{The integral is split into three different parts: $\int_0^\Lambda=\int_0^\kappa+\int_\kappa^{\kappa'}+\int_{\kappa'}^\Lambda$, with $\kappa\ll\kappa'\ll\Lambda$. The IR and UV contributions are calculated by expanding the Hankel function for small and large arguments, respectively. Furthermore, the assumption that the mass is small sets the order $\nu$ to be $\nu=3/2-\epsilon$ with $\epsilon\ll1$. This allows for an expansion in $\epsilon\approx m^2/3H^2$ in all integrals, which for the middle integral, $\int_\kappa^{\kappa'}$, simplifies to setting $\nu=3/2$.}
\item{A change of variables is performed from comoving momentum $k$ to physical momentum $p=k/a$. One then regularizes the integrals with cut-offs in the physical momentum $p$, since this is the choice that respects de Sitter symmetry.}
\end{itemize}
After implementing this procedure, we find for $m^2>0$
\begin{equation}
\frac{m^2}{g_4}=\frac{\mu^2}{g_4}+(H\tau)^2\bar\sigma^2+\frac{1}{8\pi^2}\left[\Lambda^2+2 H^2 \log\left(\frac{\Lambda}{H}\right)\right]+\frac{H^2}{8\pi^2}\left(2\gamma_{\rm E}-4+2\log2+\frac{3H^2}{m^2}\right)-\frac{m^2}{8\pi^2}\log\left(\frac{\Lambda}{H}\right)\,,
\end{equation}
in which $\Lambda$ is the UV cut-off in the physical momentum and $\gamma_{\rm E}$ is the Euler--Mascheroni constant. The divergences are renormalized through
\begin{equation}
\frac{1}{g^R_4}=\frac{1}{g_4}+\frac{1}{8\pi^2}\log\left(\frac{\Lambda}{H}\right)\,,\ \ \ \ \frac{\mu^2_R}{g^R_4}=\frac{\mu^2}{g_4}+\frac{1}{8\pi^2}\left[\Lambda^2+2 H^2 \log\left(\frac{\Lambda}{H}\right)\right]\,,
\end{equation}
resulting in
\begin{equation}
\label{unqm}
m^2=\mu^2_R+g_4^R(H\tau)^2\bar\sigma^2+g_4^R\frac{H^2}{8\pi^2}\left(2\gamma_{\rm E}-4+2\log2+\frac{3H^2}{m^2}\right).
\end{equation}
This can easily be solved for $m^2$, and one finds solutions which are strictly positive, even when $\mu^2_R\leq0$. This fact is usually referred to as radiative symmetry restoration \cite{Serreau:2011fu}, since the curved spacetime and the interactions forbid the $O(N)$ symmetry of the system from being spontaneously broken. This might not seem surprising given the initial assumption that $m^2>0$, but the existence of positive mass squared solutions is non-trivial when $\mu^2_R\leq0$. Solutions with negative $m^2$ also exist but, in those cases, the two-point function diverges in the IR, giving unphysical results.

In the next sections we will introduce a quench into the dynamics. While this will slightly alter the procedure, the main objective remains the solution of the self-consistent mass equation \eqref{selfconsdS} derived above.

\section{Quantum quenches in de Sitter}
\label{sec:quenches}

As mentioned above, a quench is defined as an instantaneous change in the parameters of a model. In the case under study, that corresponds to a change in the mass parameter, $\mu^2$, and coupling, $g_4$, of the scalar field system. We believe these quenches can arise for a number of different reasons.

In previous studies in de Sitter spacetime~\cite{Boyanovsky:1996rw,Boyanovsky:1997cr,Boyanovsky:1997xt}, the swiftness of the transition is justified by an abrupt change in the temperature of the system, which induces a sudden change in the model parameters. In the context of primordial features, however, one would expect these transitions to be due to the specific form of the scalar potential. Ref.~\cite{Joy:2007na} studies a particular example, in which an interaction between the fields prompts a fast change in the effective mass parameter of the inflaton. The motivation for the present work is the study of similar situations by using the quench approximation. In this work, however, we do not investigate the origin of quenches and they should not depend on specific details of the transitions. Therefore, this work could be applied more generally than to the study of primordial features.

Our starting point assumes exact de Sitter and negligible backreaction of the quantum fluctuations of our system in the background evolution. Furthermore, we assume the system to be in an $O(N)$ invariant state and thus we set $\bar\sigma=0$, except in the discussion of section \ref{sec:Negative}. This implies that we also do not treat the background evolution of the inflaton. All these contributions would require a fully numerical approach, which we leave for future work. Here we focus on investigating the time evolution of the effective mass as well as its asymptotic behavior. This provides a full description of the system and allows one to study different problems, such as the stationarity of the system at late times and compare it to flat spacetime case, as studied by Sotiriadis and Cardy \cite{Sotiriadis:2010si}. In that case, the system becomes stationary very soon after the quench, but in the cosmological setting of the de Sitter spacetime, it is possible, in principle, that the contributions to the effective mass vary in time in an unstable way. This is something we investigate in the following sections.

\subsection{Setup}

In order to study the quench, we define an initial state in the pre-quench stage, which is usually taken to be the ground state of the system prior to the quench. Here, we choose exactly that and assume the initial state is the Bunch--Davies vacuum $\left|0\right\rangle_{BD}$. This state is parametrized by the mass before the quench, $\mu_0$. After the quench, the Hamiltonian of the system changes, and hence the initial state is typically now an excited state of the new Hamiltonian. In particular, as will be clear below, the state will be non-Bunch--Davies with respect to the post-quench Hamiltonian.

As the quench happens, the equations the field operator obeys change, due to the change of the parameters themselves. Given that we assume that change to be instantaneous, both the value and first derivative of the field should be continuous across the quench. This implies that at conformal time $\tau_0$, when the quench happens, we have
\begin{align}
\chi^{(\nu_1)}(\tau_0,\vec{x})=\chi^{(\nu_2)}(\tau_0,\vec{x})\,,\\
\frac{d}{d\tau}\chi^{(\nu_1)}(\tau_0,\vec{x})=\frac{d}{d\tau}\chi^{(\nu_2)}(\tau_0,\vec{x})\,,
\end{align}
where the fields have been labeled with $\nu_i$ to emphasize that a set of parameters has changed. Since the initial state $\left|0\right\rangle_{BD}$ is no longer the lowest energy state of the system after the quench, one can therefore define a new vacuum and its corresponding creation and annihilation operators, $b_k^\dagger$ and $b_k$, respectively. Hence, the field is now expanded as
\begin{equation}
\chi^{(\nu_2)}(\tau,\vec{x})=\int{\frac{\text{d}^{d-1}k}{(2 \pi)^{d-1}}\left(b_k u_k^{(\nu_2)}(\tau) e^{i\vec{k}\cdot\vec{x}}+b_k^\dagger (u_k^{(\nu_2)}(\tau))^* e^{-i\vec{k}\cdot\vec{x}}\right)}\,.
\end{equation}
The constraints at $\tau_0$ given above can then be solved by a Bogoliubov transformation\footnote{Equivalently, one could keep the same expansion in $a_k^\dagger$ and $a_k$ and impose the continuity conditions on the wave-function appearing in front. Such wave-functions would be different from $u_k^{(\nu_2)}(\tau)$ and can be derived from the Bogoliubov transformation.}, which is given by
\begin{equation}
b_k=C_k a_{k}+D_k a_{-k}^\dagger\,,
\end{equation}
with
\begin{equation}
C_k=\frac{W\left((u^{(\nu_2)}_k)^*,u^{(\nu_1)}_k\right)}{W\left((u^{(\nu_2)}_k)^*,u^{(\nu_2)}_k\right)}\,,\ \ \ \
D_k=\frac{W\left((u^{(\nu_2)}_k)^*,(u^{(\nu_1)}_k)^*\right)}{W\left((u^{(\nu_2)}_k)^*,u^{(\nu_2)}_k\right)}\,,
\end{equation}
where all the wave-functions are evaluated at $\tau_0$ and $W(f,g)$ is the Wronskian, defined by
\begin{equation}
W(f,g) \equiv \frac{df}{d\tau}g-f\frac{dg}{d\tau}\,.
\end{equation}
It is straightforward to check that should the quench not occur (i.e. if $\nu_1=\nu_2$), one finds $C_k=1$ and $D_k=0$, as expected.

Given the decomposition above, it is now possible to compute the equal-time two-point correlator of the field $\chi$ after the quench. As was discussed in the previous section, this is the quantity which is required for solving the self-consistent mass equation, Eq.~\eqref{selfconsdS}, and it is also that which is observationally constrained. It is given by
\begin{align}
&_{BD}\<0|\chi(\tau_a,\vec{x})\chi(\tau_b,\vec{y})|0\>_{BD}=\nonumber\\
&\int\frac{\text{d}^{d}k}{(2 \pi)^d}e^{i\vec{k}\cdot(\vec{x}-\vec{y})}\left[C_k D_k u_k^{(\nu_2)}(\tau_a)u_k^{(\nu_2)}(\tau_b)+C_k^* D_k^* u_k^{(\nu_2)*}(\tau_a)u_k^{(\nu_2)*}(\tau_b)+\right.\nonumber\\
&\left.+\left|D_k\right|^2\left(u_k^{(\nu_2)*}(\tau_a)u_k^{(\nu_2)}(\tau_b)+u_k^{(\nu_2)}(\tau_a)u_k^{(\nu_2)*}(\tau_b)\right)+u_k^{(\nu_2)}(\tau_a)u_k^{(\nu_2)*}(\tau_b)\right]\,.\label{bogtrans}
\end{align}
Again, it is clear that in the absence of the quench only the last term survives, which is the result shown in Eq. \eqref{twopointunquench}.

The sections that follow will be dedicated to performing the calculations for different scenarios. For the simplest cases we are able to use analytical methods, which give a general picture of the results. We then complement those estimates with numerical calculations of the time evolution of the mass and interpret the results.

\subsection{Analytical estimates}
\label{sec:Analytical}

Before presenting our results, we make a note of difficulties we encounter and the simplifying assumptions we use in order to make the problem analytically tractable.
As was mentioned above, the state after the quench is no longer the Bunch--Davies vacuum of the system. Therefore, de Sitter invariance is broken and the two-point function of $\phi$ is no longer time-independent, in general. The first approximation we make is related to that: we will assume that time dependence to be negligible, at least in what concerns its effect on the two-point function. By this we mean that we calculate the two-point function assuming the wave-functions, $u_k^{(\nu_i)}$, to be the solutions from the unquenched case (i.e. with constant mass), as given by Eq.~\eqref{wavfunc}. This approximation is necessary given that it is impossible to (analytically) solve the Klein--Gordon equation for a general time-varying mass. Furthermore, as mentioned above, it has been shown that this is a very good approximation in flat spacetime~\cite{Sotiriadis:2010si}, and hence this is a justified approach.

Another difficulty that arises is the calculation of the integral of the power spectrum. It will generally involve integrating four Hankel functions with different arguments, which cannot be done analytically unless the order of the Hankel functions is a half integer. For this reason, we only treat masses close to $0$ or $\sqrt2 H$, due to the simplicity of the corresponding Hankel functions of orders $3/2$ and $1/2$, respectively. This means that, in some cases, we do not explicitly solve the self-consistent mass equation, but instead check if certain transitions are possible and focus on closed form formulae. This does not undermine the generality of the results, although it makes the physical interpretation more transparent. Note, however, that this care is not necessary in flat spacetime, given the analytical simplicity of the wave-functions.

To overcome this, we employ the same procedure as in section \ref{NoQuench}, by splitting the momentum integral into three parts, which we call the IR, middle and UV integrals. We also change variables to physical momentum, so that UV cut-offs are correctly defined. UV contributions are rather simple to evaluate---they turn out to be the same as in the unquenched case, with the mass $m$ substituted by the mass after the quench.\footnote{This is strictly true only for $\tau>\tau_0$. At the instant in which the quench happens, $\tau=\tau_0$, the continuity of the two-point function implies that the UV contributions are still dependent on the mass before the quench. We disregard that point in time in all calculations.} This is not surprising, as the UV limit should not depend on initial conditions whichever they may be. The UV contribution to the self-consistent mass equation is therefore given by
\begin{equation}
\frac{m^2}{g_4}\supseteq \frac{1}{8\pi^2}\left[\Lambda^2+\left(2 H^2-m^2\right) \log\left(\frac{\Lambda}{H}\right)\right]\,,
\end{equation}
where $m$ denotes again the effective mass after the quench. Renormalization is performed in the same way as in the unquenched case.

\subsubsection{Asymptotic mass}

The first calculation we perform is the limit $x=\tau/\tau_0\rightarrow0$ of the self-consistent mass equation. The mass after the quench is now:
\begin{equation}
m^2_\infty=\mu^2_R+g_4^R(-H\tau)^{2}\int{\frac{\text{d}^{3}k}{(2\pi)^{3}}\frac{\pi}{4}(-\tau)\left|H^{(1)}_{\nu_2^\infty}(-k\tau)\right|^2 }\,,
\end{equation}
where we have also set $\bar\sigma$ to $0$. The integral can actually be calculated without approximations so that the result becomes
\begin{equation}
\label{asympmass}
m^2_\infty=\mu^2_R+\frac{g_4^RH^2}{16 \pi^2} \left(\frac{m^2_\infty}{H^2}-2\right)\left[\log 4 -1-\Psi\left(\nu_2^\infty-1/2\right)-\Psi\left(-\nu_2^\infty-1/2\right)\right]\,,
\end{equation}
where $\Psi(x)$ is the Digamma function, defined as the logarithmic derivative of the Gamma function, $\Psi(x)\equiv\Gamma'(x)/\Gamma(x)$. This result can now be approximated for masses close to $0$ and one would find the same result as in the unquenched case, Eq.~\eqref{unqm}. The point to note in this result is how different it is from the flat spacetime case, in which the system retains some memory of its state before the quench, even in the asymptotic late time limit. As shown in Ref.~\cite{Hung:2012zr}, the asymptotic mass is a function of the pre-quench mass, $\mu_0$. That does not seem to happen in de Sitter spacetime, given that Eq.~\eqref{asympmass} is independent of the original mass. This is related to the evolution of the cosmological horizon. As is well known, scales $k^{-1}$ larger than the comoving horizon size $(aH)^{-1}$ are enhanced in an accelerating spacetime. These IR scales are the ones that end up dominating the calculation of the two-point function. Given that the horizon shrinks with time, the number of super-horizon scales increases with time. In the presence of a quench, however, the number of scales that exited the horizon before the quench is constant, while the number of modes that are enhanced after the quench increases indefinitely. After sufficient time, the contribution to the integral of the propagator from pre-quench modes becomes negligible in comparison to the scales that became super-horizon after the quench. As a consequence, the dependence of the effective mass on the pre-quench parameters disappears.\footnote{Note, however, that this is only true because $\mu_0^2\leq0$ is not allowed. If it were, IR divergences would appear, and thus the contribution from pre-quench modes would be non-negligible (and infinite).} These effects are not present in flat spacetime and thus the dependence on the initial mass is always present.

This result is not sufficient, on its own, without first making sure that the mass converges in general. While in the flat situation the convergence to a stationary mass is fast enough for one to assume the asymptotic result is valid shortly after the quench, the same is not clear in a curved spacetime, and that is the reason why one must find a more complete time evolution, thus checking both convergence as well as its rate of change.

Note, however, that, should the mass converge to a constant at some time, then the result above must be valid, since for a constant mass, the system is in a de Sitter invariant state, equivalent to the unquenched scenario. Hence, if we can prove that it does converge, we already have the expression for the asymptotic mass, Eq. \eqref{asympmass}.

\subsubsection{Approximate time evolution}\label{ATE}

We now move on to the time evolution. We begin by studying it for specific transitions of masses close to $0$ or $\sqrt2 H$. These cases are interesting for different reasons. Firstly, as mentioned before, they correspond to half-integer orders of the Hankel functions, which simplifies the wave-functions considerably. Furthermore, the $m\approx 0$ case is the relevant situation in inflation, since then the quantum perturbations are enhanced by the accelerated expansion. The other situation, $m=\sqrt2H$, is the conformal case, in which one can completely disregard the cosmic expansion from its evolution---its wave-functions turn out to be equal to those of the massless case in flat spacetime. Furthermore, in a de Sitter-invariant state, its mass does not receive any contributions from the interactions, as can be seen in Eq.~\eqref{asympmass}.

The other main approximation we employ here is the use of the wave-functions obtained for constant masses, i.e. , instead of solving the full equation of motion,
\begin{equation}
\label{EOMWF}
u_k''+\left[k^2+\frac{1}{\tau^2}\left(\frac{m^2(\tau)}{H^2}-2\right)\right]u_k=0\,,
\end{equation}
we solve only for $m^2(\tau)=$const. as a first approximation. This will result, in general, in a time-dependent solution of the mass equation, Eq.~\eqref{selfconsdS}, which we label $m_1(\tau)$. Ideally, one could go further in the approximation by substituting the solution $m_1(\tau)$ in the evolution equation, Eq.~\eqref{EOMWF} and thus finding the second approximation, $m_2(\tau)$, by solving the mass equation once more. Repeating this procedure should result in more and more accurate results with each iteration and convergence to the real effective mass. However, provided the difference between the first iterations is negligible, it is sufficient to use the approximation of constant mass and thus stop at $m(\tau)\approx m_1(\tau)$. We will estimate the size of that difference by comparing the solutions of Eq.~\eqref{EOMWF} for constant mass ($u_0(\tau)$) and for the first approximation $m_1(\tau)$ ($u_1(\tau)$). In particular, we calculate the error, $e_u$, with
\begin{equation}
\label{error}
e_u=\max \left|1-\frac{|u_1(\tau)|^2}{|u_0(\tau)|^2}\right|\,.
\end{equation}
Given that we expect the iterative approach to converge, this error calculation essentially gauges whether the first iteration, $m_1(\tau)$, is sufficiently accurate. An alternative to this procedure would be to check how large the time derivatives of $m_1(\tau)$ are. A particular test would be the calculation of the following derivative:\footnote{A derivation of this quantity can be made by obtaining the rate of change of the frequency, $\omega^2$ (given in square brackets in Eq.~\eqref{EOMWF}),
\begin{equation}
\frac{d(\omega^2)}{d\tau}=\frac{1}{\tau^2}\left(\frac{d(m^2(\tau)/H^2-2)}{d\tau}-\frac2\tau (m^2(\tau)/H^2-2)\right)\,,\nonumber
\end{equation}
and comparing the contribution from the time-dependent mass (the first term) to the contribution due to the time-dependent background (the second term).}
\begin{equation}
\label{fastslow}
\left|\frac12\frac{d\log\left|m^2(\tau)/H^2-2\right|}{d\log \tau}\right|\ll1\Rightarrow\left|\frac{d\log\left|m^2(\tau)/H^2-2\right|}{dt}\right|\ll 2 H\,,
\end{equation}
where $t$ is cosmic time. Note that the second inequality explicitly shows the connection of this test to the time scale of the problem, the Hubble rate, $H$, thus providing the physical interpretation to how slow the evolution needs to be for the correctness of the constant mass approximation.\footnote{Note that using the opposite inequality in Eq.~\eqref{fastslow} would correspond to the quench itself, in which the transition happens in a much shorter time-scale than $H^{-1}$.} While being more physically intuitive, this method is less accurate in predicting whether the first iteration is sufficiently good, which is why we use the expression given in Eq.~\eqref{error} to estimate the error. 

In the calculations that follow, we begin by assuming the corrections are small, similarly to what occurs under an adiabatic approximation, in which one assumes the evolution of the mass to be slow enough for it not to affect the equations of motion substantially. We will revisit the accuracy of this approximation in section \ref{sec:Numerical}, thereby justifying our approach.
\\

\noindent\textbf{Transition 1:} $\mu_0\approx 0\ \rightarrow\ m=\sqrt2 H$
\\

The first case we will consider is the transition from $\mu_0\approx 0$ to $m=\sqrt2 H$. By $\mu_0\approx 0$, we mean we use the same approximations as in the unquenched case, i.e. the order of the Hankel function before the quench is $\nu_1=3/2-\epsilon_1$ with $\epsilon_1\ll1$ and we expand in powers of $\epsilon_1\approx \mu_0^2/3H^2$. At lowest order in $\epsilon_1$, we find
\begin{gather}
\label{Res3212}
2H^2=m^2=\mu_R^2+\frac{g_4^R H^2}{8\pi^2}x^2\left[\left(\frac{1}{\epsilon_1}-3-2 \log(1-x)\right)(x-2)^2-1\right]\,,
\end{gather}
in which $x=\tau/\tau_0$. We can see that this result does converge to a constant at late times ($x\rightarrow 0$), and becomes $m^2=\mu_R^2$, in agreement with our estimate from Eq.~\eqref{asympmass}.

The conclusion seems to be that should we have $\mu_R^2=2H^2$, a transition does exist from $\mu_0\ll H$ to $m\approx\sqrt2 H$, given that the time evolving part is very small, when compared to $2H^2$. Should that not be the case, not only is it not guaranteed that the evolution is slow enough, but the result is not even consistent with the original assumption. Recall that we are checking whether the transition exists by assuming the final mass is $m=\sqrt2 H$ and attempting to find parameters $\mu_R^2$, $g_4^R$ and $\epsilon_1$ for which the solution is consistent. If we find the time dependent part to be very large, consistency is violated and our result for the two-point function could no longer be valid. We check this in section \ref{sec:Numerical} using numerical calculations and find no such problems.
\\

\noindent\textbf{Transition 2:} $\mu_0=\sqrt2 H\ \rightarrow\ m\approx 0$
\\

We now look into the inverse transition, $\mu_0=\sqrt2 H\ \rightarrow\ m\approx 0$. We use the same approximations as in the previous case, but expand now in $\epsilon_2\approx m^2/3H^2$. Again, at first order in this parameter, we find
\begin{equation}
\label{q32p}
m^2=\mu_R^2+\frac{g_4^RH^2}{16\pi^2}\left[4\gamma_{\rm E}-5+\log 16 +4x+x^4+4 \log(1-x)-4\log x\right]\,.
\end{equation}
This result does not match our original predictions for the final masses, due to an apparent divergence when $x\rightarrow 0$. This is re-analyzed in section \ref{sec:Numerical}, and the numerical results show no divergences, indicating that this is a problem owing to the expansion in $\epsilon$.
\\

\noindent\textbf{Transition 3:} $\mu_0\approx 0\ \rightarrow\ m\approx 0$
\\

The final case we deal with here is the transition $\mu_0\approx 0\ \rightarrow\ m\approx 0$, now expanded both in $\epsilon_1\approx \mu_0^2/3H^2$ and $\epsilon_2\approx m^2/3H^2$. The self consistency condition for this case is
\begin{equation}
\label{q3232}
m^2=\mu_R^2+\frac{g_4^RH^2}{8\pi^2}\left[-4+2\gamma_{\rm E}+2\log 2+\frac{1}{\epsilon_1}\left(1+(\epsilon_2-\epsilon_1)\left(\frac23-\frac23x^3+2\log x\right)\right)\right]\,.
\end{equation}
We can see that the late time limit ($x\rightarrow0$) again results in a divergence, unless there is no quench, i.e. $\epsilon_2=\epsilon_1$. The logarithmic divergences are now slightly more complicated, with one term being identical to that of \textbf{transition 2}, while the other is dependent on $\epsilon_2$. Again, for this case, it will be made clear in the next section that the problem comes from the expansion in $\epsilon_1$ and $\epsilon_2$, rather than being symptomatic of a ``dynamical impossibility".

\subsection{Numerical and re-summed results}
\label{sec:Numerical}

In this section we perform the calculations from the previous section again but using numerical techniques. Instead, this allows one to see that the full results from the previous calculations do now match the final mass estimates from Eq.~\eqref{asympmass} once we implement a re-summation technique and that most of the other issues are solved. However, we do still use the same approximation, in which we take the mass to be constant for the purposes of calculating the integrals. We remind the reader that we have defined the parameters $\epsilon_1$ and $\epsilon_2$ as
\begin{equation}
\label{epsilons}
\epsilon_1 \equiv \frac32-\sqrt{\frac94-\frac{\mu_0^2}{H^2}}\approx \frac{\mu_0^2}{3H^2}\ \ \textrm{and} \
\ \epsilon_2 \equiv \frac32-\sqrt{\frac94-\frac{m^2}{H^2}}\approx \frac{m^2}{3H^2}\, ,
\end{equation}
respectively. Recall as well that conformal time is defined in the range $-\infty<\tau<0$, so that $x=\tau/\tau_0$ is positive and approaches $x\rightarrow 0$ in the far future.
\\

\noindent\textbf{Transition 1:} $\mu_0\approx 0\ \rightarrow\ m=\sqrt2 H$
\\

\begin{figure}[h]
    \centering
    \includegraphics[scale=0.8]{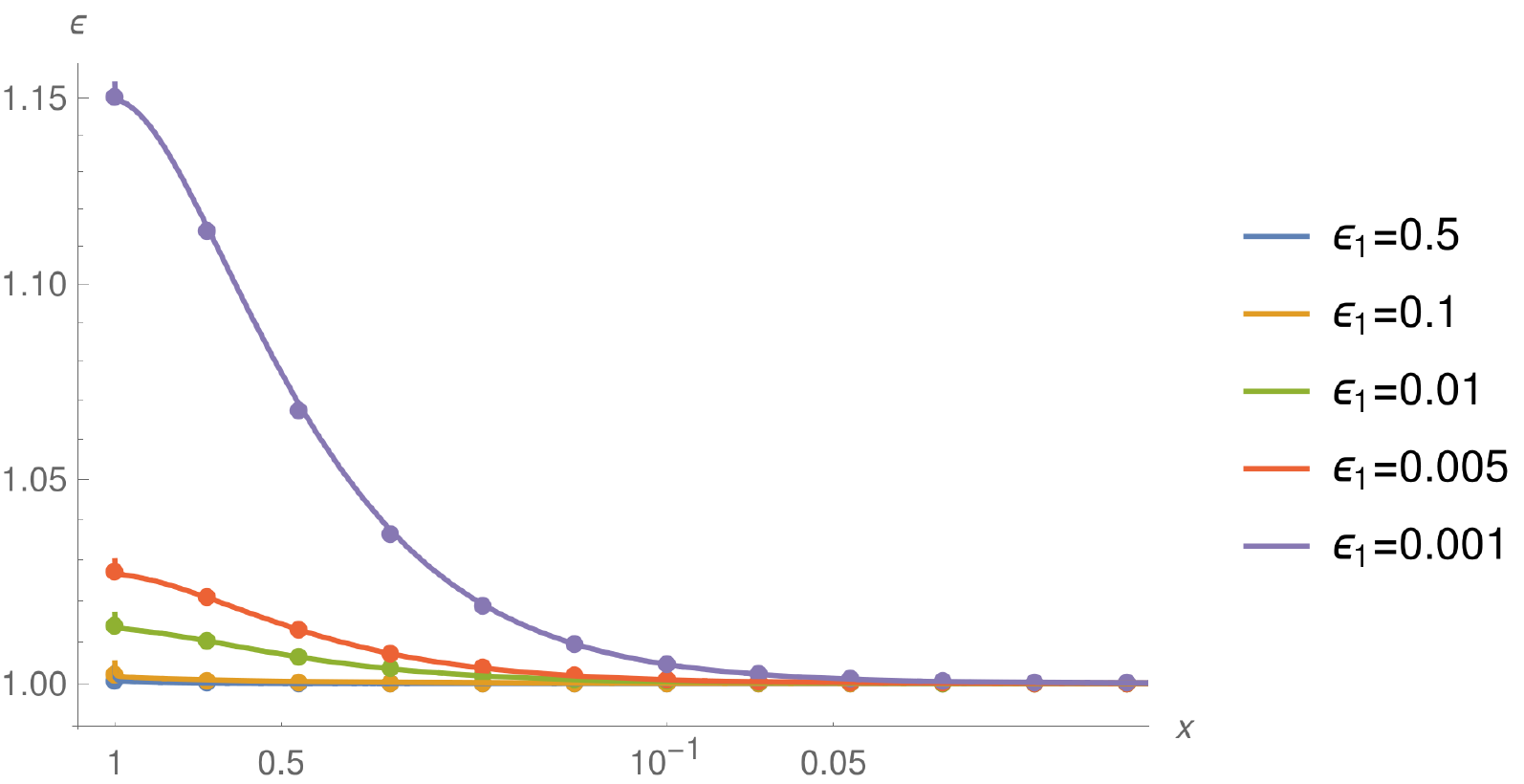}
    \caption{Evolution of $\epsilon_2(x)$ for transition 1 (dotted) as compared to the analytical result (solid) for $g_4^R=0.01$, varying $\epsilon_1$. }
    \label{fig3212e1}
\end{figure}

\begin{figure}[h]
    \centering
    \includegraphics[scale=0.8]{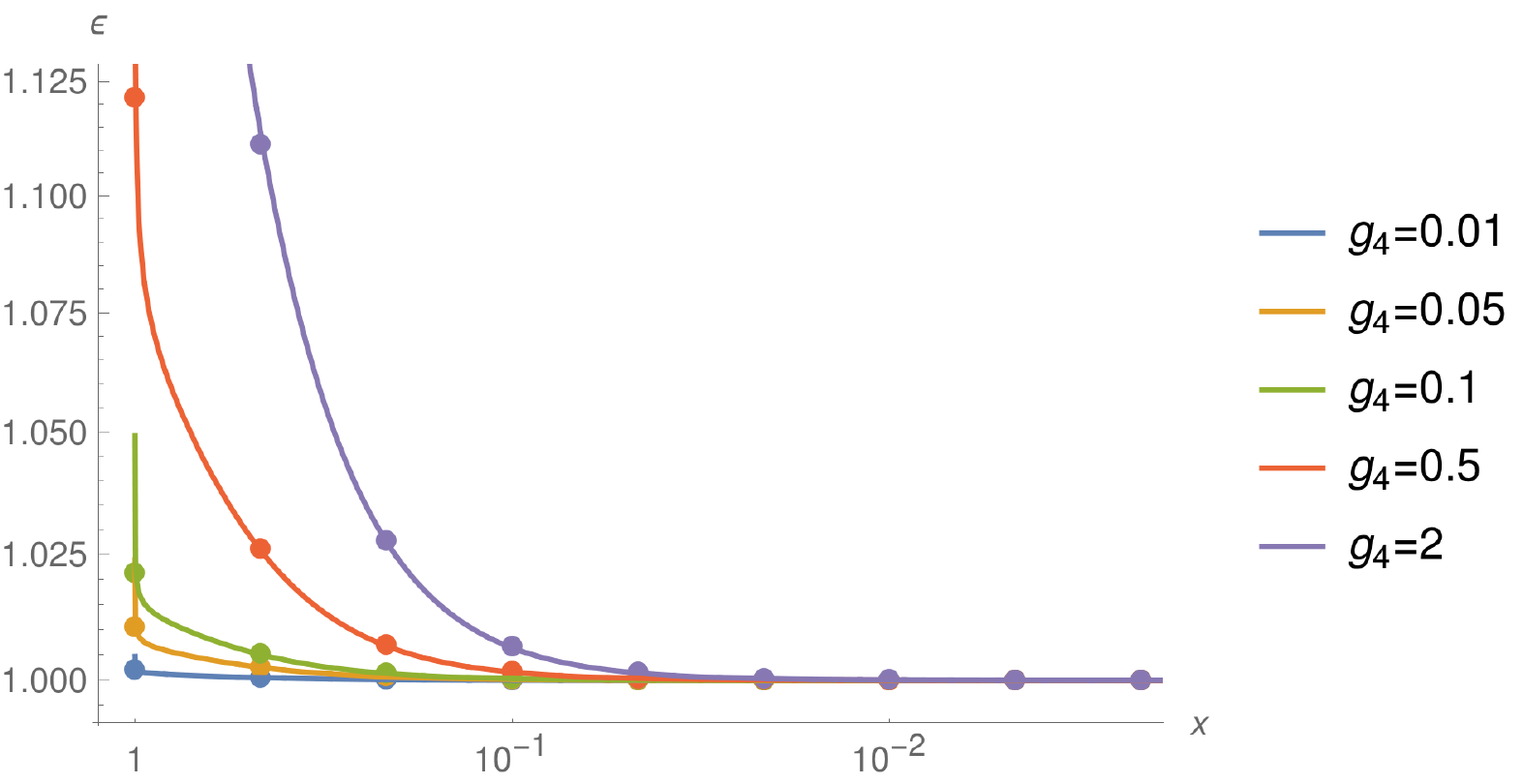}
    \caption{Evolution of $\epsilon_2(x)$ for transition 1 (dotted) as compared to the analytical result (solid) for $\epsilon_1=0.1$, varying $g_4^R$. }
    \label{fig3212g4}
\end{figure}

Let us follow the same order as before and start with the case $\mu_0\approx 0\rightarrow m=\sqrt2 H$. We have seen that, in order for this transition to occur, one must have $\mu_R^2=2H^2$, so we choose that value for the mass parameter. We demonstrate the dependence on the remaining parameters by plotting $\epsilon_2$ as a function of $x=\tau/\tau_0$ for different values of the original mass, $\mu_0$ (labeled by $\epsilon_1$), and the coupling strength, $g_4^R$ in two different plots, in Figs. \ref{fig3212e1} and \ref{fig3212g4}.

Firstly, we notice that the analytical expression obtained above in Eq.~\eqref{Res3212} is a very good approximation to the numerical solution in all situations and for all values of $x$. This is somewhat surprising, given that that expression was derived for a specific final mass. Furthermore, from Fig. \ref{fig3212e1}, we see that even when $\epsilon_1$ is not so small, as exemplified by the case $\epsilon_1=0.5$, our original approximation almost reproduces the numerical results, with only a small deviation of less than $0.01\%$ around $x=1/3$. It would fail for larger values of $\epsilon_1$, but those cases are somewhat less interesting, since the initial and final masses are too similar.

As expected, evolution is faster and more pronounced in the cases in which the coupling strength, $g_4^R$, is larger. The dependence on the initial mass, $\epsilon_1$, seems to indicate that there is less evolution for larger initial masses, which is to be expected given the terms with $H/\mu_0$ present in Eq.~\eqref{Res3212}.
\\\\\\

\noindent\textbf{Transition 2:} $\mu_0=\sqrt2 H\ \rightarrow\ m\approx 0$
\\

\begin{figure}[h]
    \centering
    \includegraphics[scale=0.8]{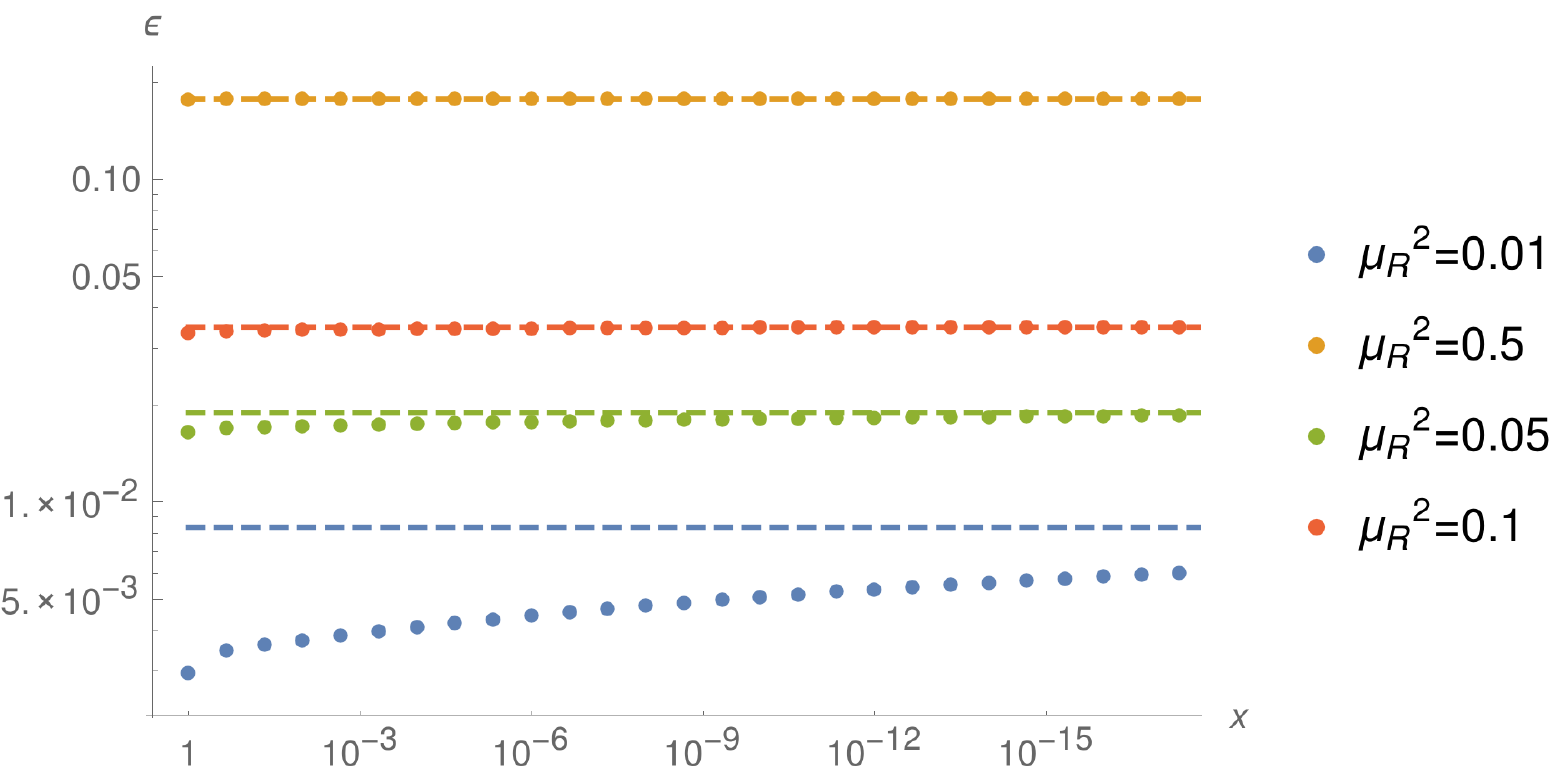}
    \caption{Numerical evolution of $\epsilon_2(x)$ for transition 2 (dotted), showing the asymptotic mass (dashed) for $g_4^R=0.01$, varying $\mu_R^2$ (shown in units of $H^2$). }
    \label{fig1232g4}
\end{figure}

Moving now to the results for the inverse transition, $\mu_0=\sqrt2 H\rightarrow m\approx 0$, we are interested again in showing that this transition is possible under our approximations. Our analytical result from the previous section hinted at convergence problems in the late time limit, and here we check whether those issues are present when \emph{no expansion} in $\epsilon_2$ is made. Given that we are checking \textbf{transition 2}, we set the initial mass to $\mu_0=\sqrt2 H$, or equivalently $\epsilon_1=1$. We begin by showing the results for $\epsilon_2$ by varying the mass parameter, $\mu_R^2$, in Fig.~\ref{fig1232g4}. We also plot the asymptotic value (dashed curve) as predicted by Eq.~\eqref{asympmass}.

We note that convergence is indeed achieved and that it agrees with the expectation for the asymptotic mass from Eq.~\eqref{asympmass}. Furthermore, we note that in the analytical result for the evolution, Eq.~\eqref{q32p}, the r.h.s. did not depend on the final mass, $m$ (or $\epsilon_2$), which would imply that the time-evolving part of the solution for $\epsilon_2$ would not change among different choices of $\mu_R^2$. It is clear from Fig.~\ref{fig1232g4}, however, that the evolution is different from case to case, which emphasizes the need for an extension to that analytical result.

It turns out that one can improve the analytical estimate substantially, by changing the divergent $\log x$ term into a dynamical renormalization group (DRG) inspired expression~\cite{Burgess:2009bs, Dias:2012qy}. The resulting mass equation becomes
\begin{equation}
\label{mass1232corr}
m^2=\mu_R^2+\frac{g_4^RH^2}{16\pi^2}\left[4\gamma_{\rm E}-5+\log 16 +4x+x^4+4 \log(1-x)+\frac{2}{\epsilon_2} \left(1-x^{2\epsilon_2}\right) e^{-\frac{3\epsilon_2}{2}}\right],
\end{equation}
where the last term has been added. It is easy to show that this term is equal to $-4 \log x$ in the limit $\epsilon_2\rightarrow 0$, as required. Given the similarity with the DRG method, we also call this new expression the re-summation of the previous one, given that one understands this correction as the sum of infinite terms with different powers of $\log x$.\footnote{A similar problem was detected in scattering calculations in kinematic regions where there is a large hierarchy of scales, the so-called Sudakov region~\cite{Sudakov:1954sw}, for which the Kinoshita--Lee--Nauenberg theorem \cite{Kinoshita:1962ur,Lee:1964is} is not valid. Re-summation of the large logarithms that appear is then required to make sense of the result. The techniques used for that case offered inspiration to the solution to very similar problems in inflationary correlation function calculations~\cite{Dias:2012qy} dealing with secular divergences \cite{Seery:2010kh}. The logarithms that appear in the present work are also, in fact, due an IR divergence arising because of the evolution of the system towards a massless state. After re-summation, it is clear that the presence of a finite mass resolves the divergence.}

\begin{figure}[h]
    \centering
    \includegraphics[scale=0.8]{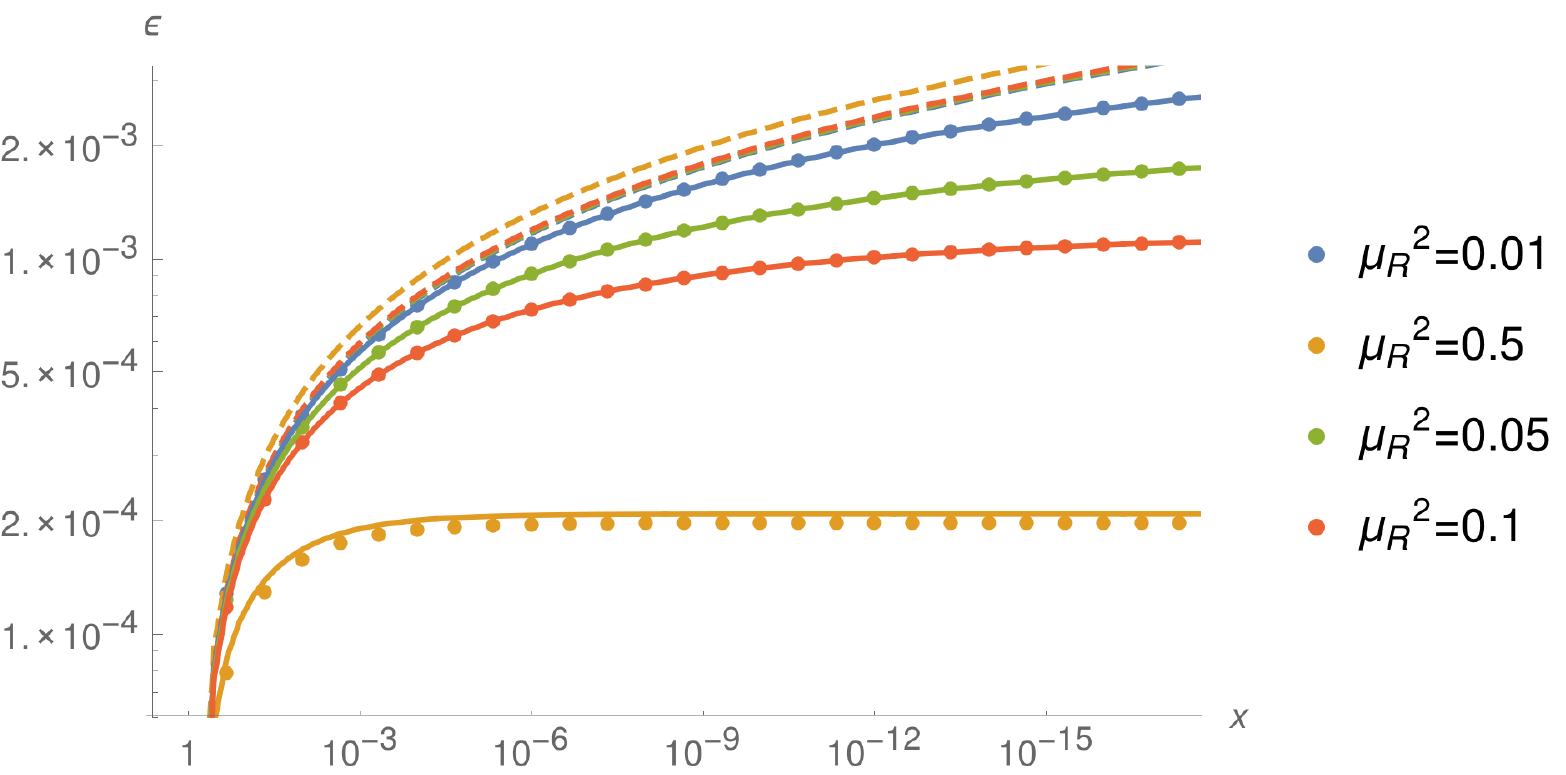}
    \caption{Numerical evolution of $\epsilon_2(x)$ for transition 2 (dotted) as compared to both the corrected (solid) and uncorrected (dashed) analytical results, for $g_4^R=0.01$, varying $\mu_R^2$ (shown in units of $H^2$) and rescaled by $\mu_R^2$. }
    \label{fig1232g4Re}
\end{figure}

The improvement the re-summation brings to the result can be seen in the plot of Fig.~\ref{fig1232g4Re}, in which the results have been rescaled according to $\mu_R^2$ and we plot both the numerical results and the solution to the new mass equation, Eq.~\eqref{mass1232corr}.

The uncorrected result of Eq.~\eqref{q32p} is also shown in dashed lines. In spite of there being a substantial improvement, there is still a visible discrepancy for the case with the higher mass. This is expected, as the analytical result was derived for small masses, $m^2\ll H^2$ and the heavier example is already at $m^2\approx H^2/2$.

\begin{figure}[h]
    \centering
    \includegraphics[scale=0.8]{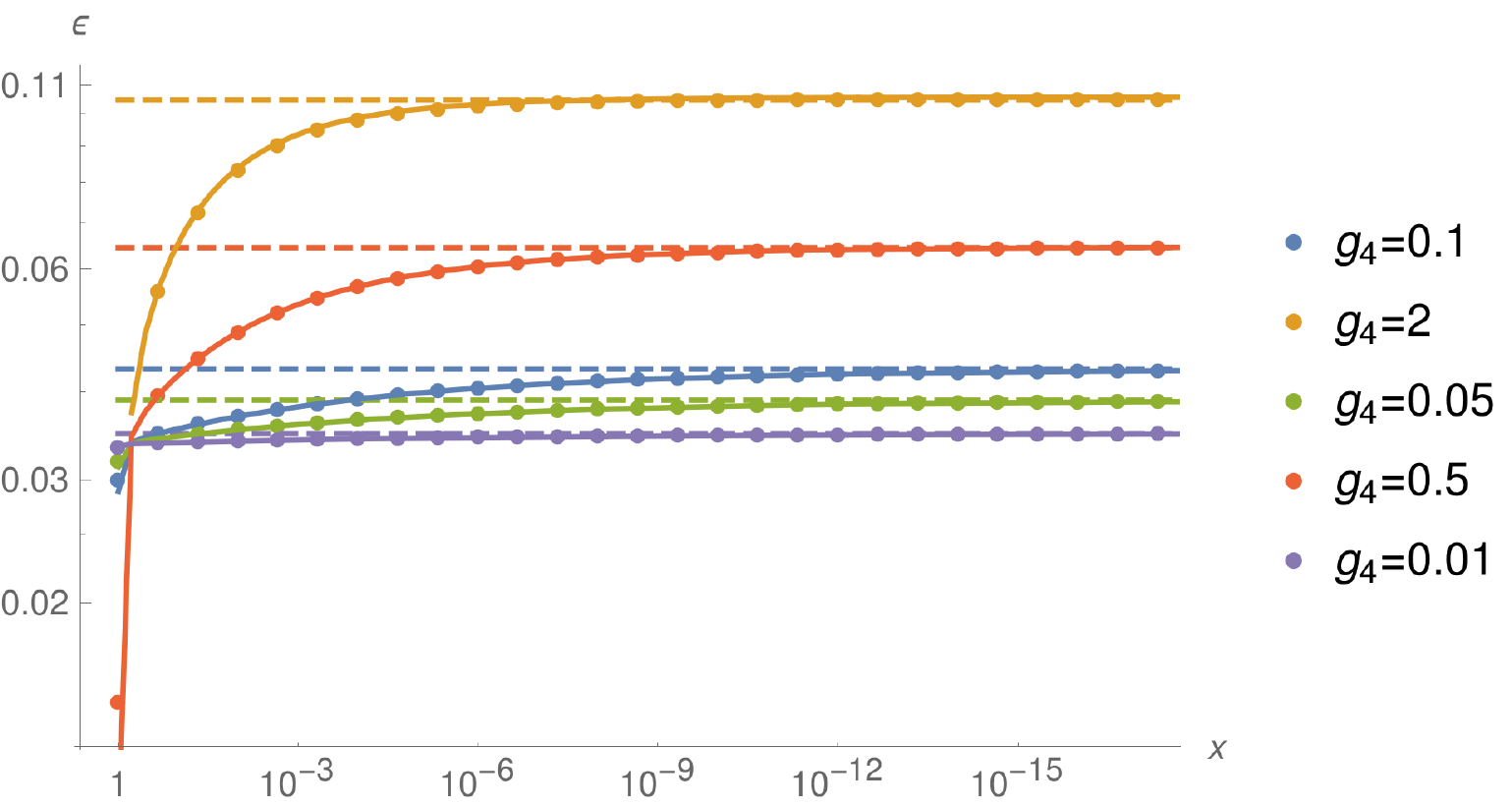}
    \caption{Numerical evolution of $\epsilon_2(x)$ for transition 2 (dotted) as compared to the corrected analytical results (solid) and showing the asymptotic mass (dashed), with $\mu_R^2/H^2=0.1$, varying $g^R_4$. }
    \label{fig1232mu}
\end{figure}

All the cases presented in Figs.~\ref{fig1232g4} and \ref{fig1232g4Re} have $g_4^R=0.01$ and the contribution from the time evolution parts to the final result was not very large. The results presented in Fig.~\ref{fig1232mu} show the dependence on $g_4^R$ for higher values of the coupling. We see that, once again, the corrected result does very well in all cases and that it converges to the asymptotic result of Eq.~\eqref{asympmass}.

We note that when $g_4^R$ becomes large, the initial evolution can become quite fast, as expected, given the effect of the interaction in Eq.~\eqref{mass1232corr}. A quick analysis of that equation shows that the evolution is slower for larger $\mu_R^2$, since in those cases the interaction terms become almost negligible in comparison to $\mu_R^2$.
\\

\noindent\textbf{Transition 3:} $\mu_0\approx 0\ \rightarrow\ m\approx 0$
\\

\begin{figure}[h]
    \centering
    \includegraphics[scale=0.8]{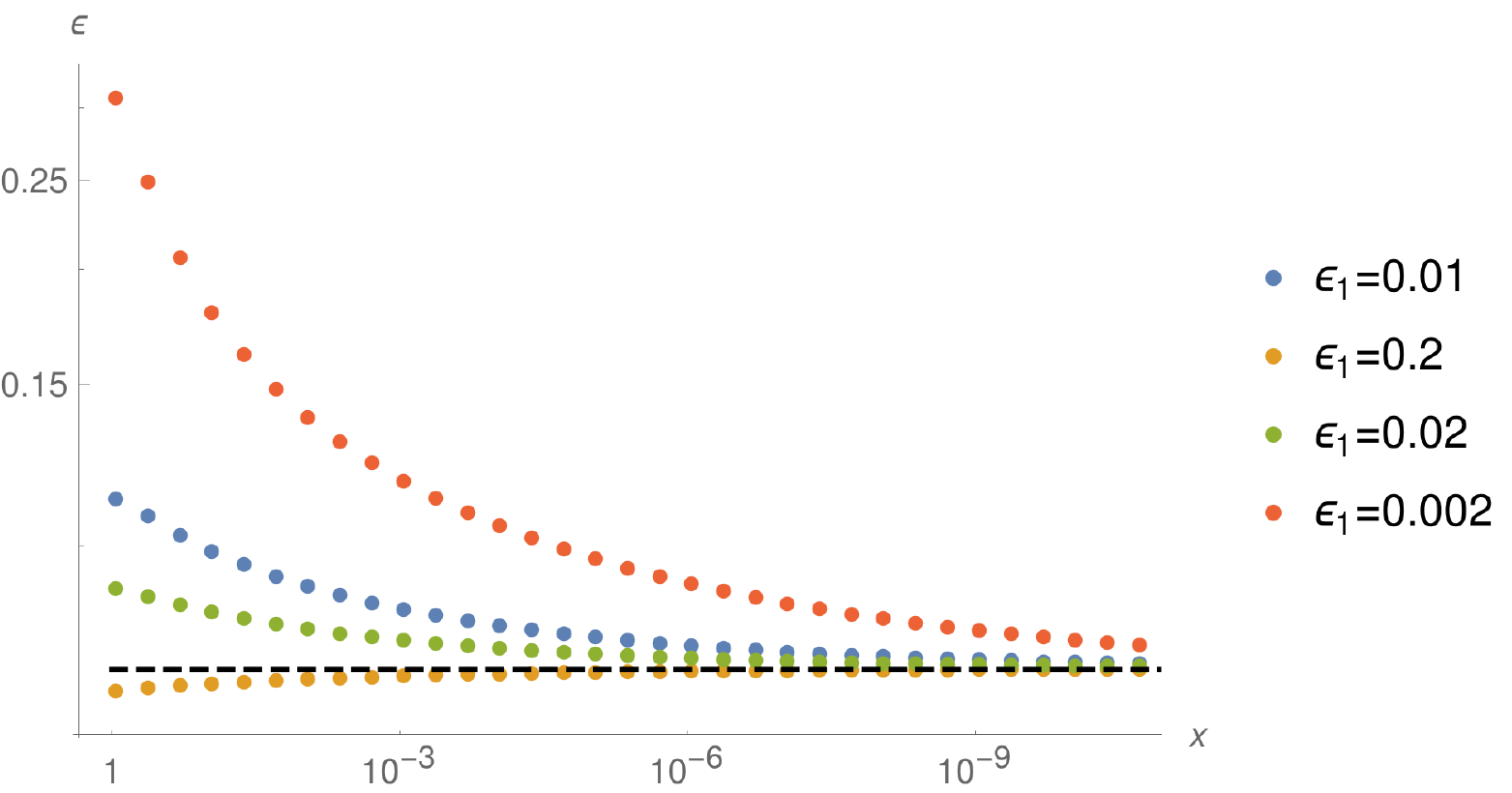}
    \caption{Numerical evolution of $\epsilon_2(x)$ for transition 3 (dotted) and showing the asymptotic mass (dashed) with $\mu_R^2/H^2=0.2$, $g^R_4=0.1$, varying $\epsilon_1$. }
    \label{fig3232e1}
\end{figure}

Let us now look at the more general case in which no mass is fixed. We focus on the cases in which the masses are small in order to compare with our results for the transition $\mu_0\approx 0\ \rightarrow\ m\approx 0$. One of the conclusions following from the expression for the asymptotic mass, Eq.~\eqref{asympmass}, was that, when $x\rightarrow 0$, the mass after the quench, $m$, should not depend on the mass before the quench, $\mu_0$. Fig.~\ref{fig3232e1} shows the time evolution of $\epsilon_2$ for the quench with parameters given by $\mu_R^2=0.2$, $g_4^R=0.1$ and varying $\epsilon_1$.

\begin{figure}[h]
    \centering
    \includegraphics[scale=0.8]{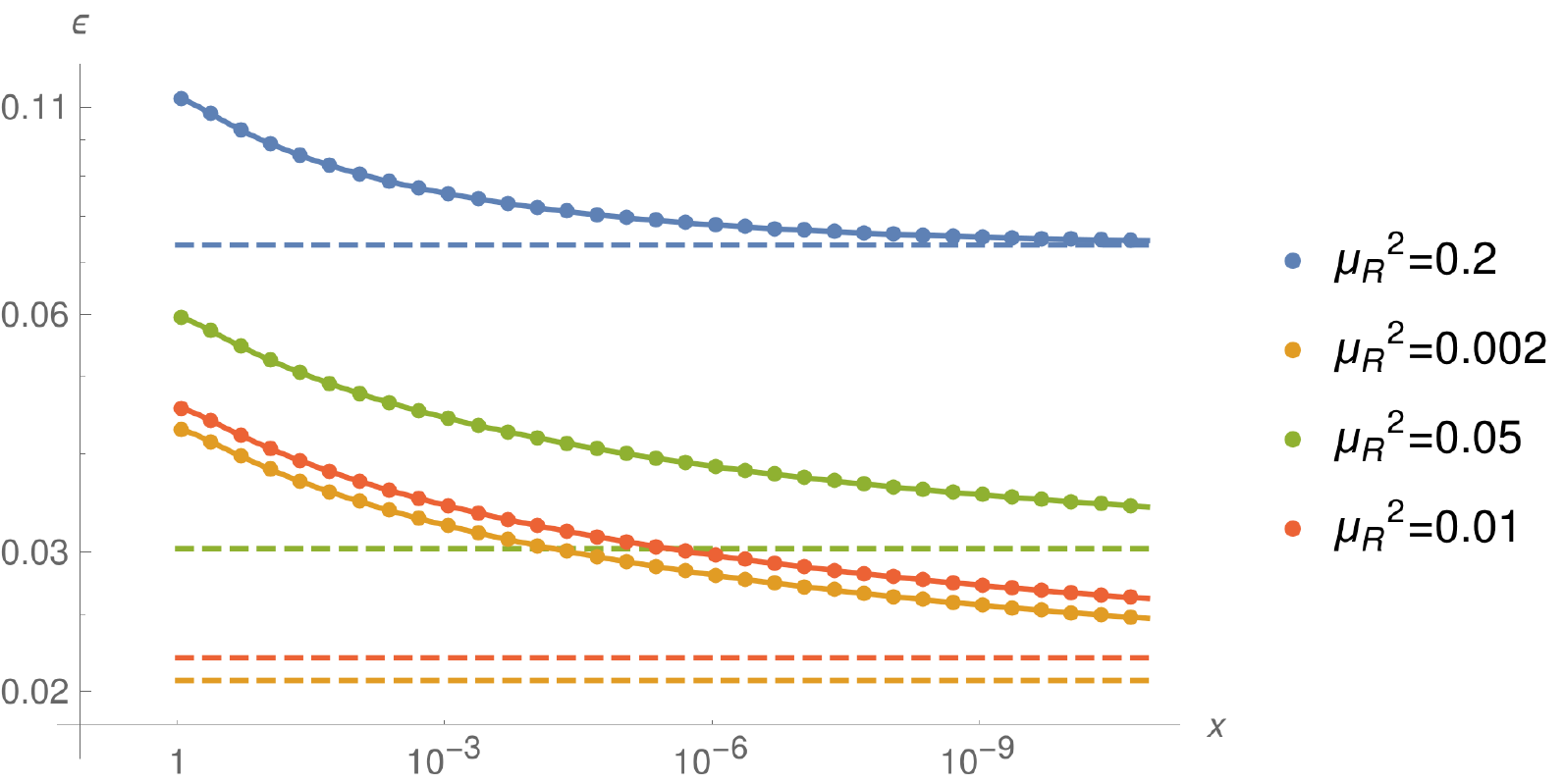}
    \caption{Numerical evolution of $\epsilon_2(x)$ for transition 3 (dotted) as compared to the corrected analytical results (solid) and showing the asymptotic mass (dashed), with $\epsilon_1=0.01$, $g^R_4=0.1$, varying $\mu_R^2$ (shown in units of $H^2$). }
    \label{fig3232mu}
\end{figure}

\begin{figure}[h]
    \centering
    \includegraphics[scale=0.8]{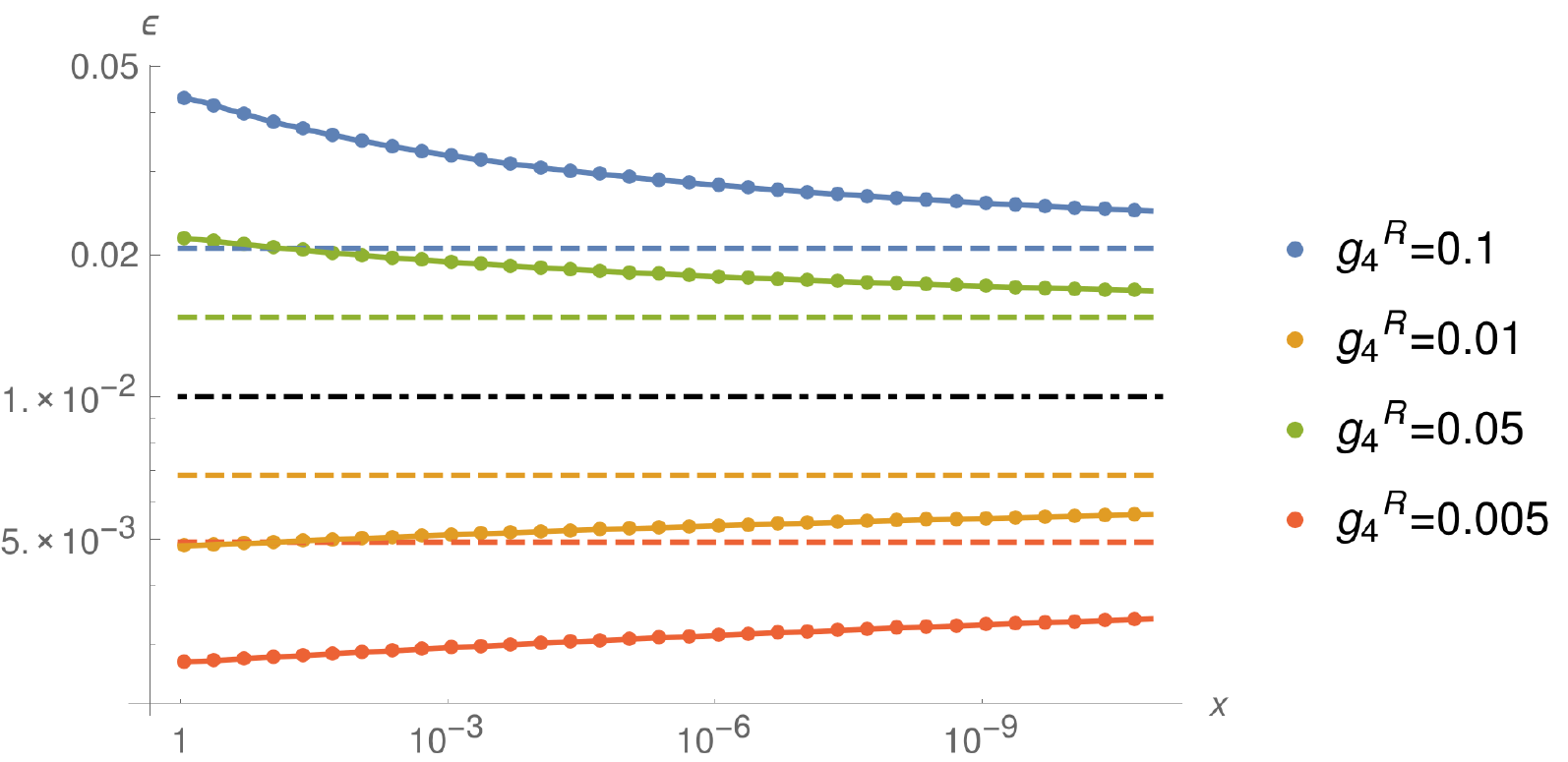}
    \caption{Numerical evolution of $\epsilon_2(x)$ for transition 3 (dotted) as compared to the corrected analytical results (solid) and showing the asymptotic mass (dashed), with $\epsilon_1=0.01$ (also shown as dot-dashed in the middle), $\mu_R^2/H^2=0.002$, varying $g^R_4$. }
    \label{fig3232g4}
\end{figure}

It is clear that, in spite of the previous analysis of Eq.~\eqref{q3232} indicating a divergent behavior at late times, the masses converge to the same value---that given by Eq.~\eqref{asympmass}. Again, in this case, it is possible to find a better approximation to the results, by drawing inspiration from dynamical renormalization group techniques \cite{Dias:2012qy,Burgess:2009bs} and applying them to Eq~\eqref{q3232}. This amounts to exponentiating the divergent terms, which results in the following expression
\begin{equation}
\label{q3232ren}
m^2=\mu_R^2+\frac{g_4^RH^2}{8\pi^2}\left[-4+2\gamma_{\rm E}+2\log 2+\frac{1}{\epsilon_2}+\frac{\epsilon_2-\epsilon_1}{\epsilon_1\epsilon_2}x^{2\epsilon_2}e^{\frac{2\epsilon_2}{3}(1-x^3)}\right]\,.
\end{equation}
It is now clear that this solution has the correct asymptotic limit up to $O(\epsilon_2)$ corrections, given by
\begin{equation}
m^2=\mu_R^2+\frac{g_4^RH^2}{8\pi^2}\left[-4+2\gamma_{\rm E}+2\log 2+\frac{1}{\epsilon_2}\right]\,.
\end{equation}
This is equivalent to the result for the unquenched situation, Eq.~\eqref{unqm}, as expected from our previous analysis. We can see that this matches the numerical results very well in the plots that follow. We show both the effect of varying $\mu_R^2/H^2$ in Fig.~\ref{fig3232mu} and the dependence on $g_4^R$ in Fig.~\ref{fig3232g4}. Again, we show that, asymptotically, there is convergence towards the values given by Eq.~\eqref{asympmass}.

We see that the difference in mass $\epsilon_2-\epsilon_1\approx (m^2-\mu_0^2)/3H^2$ is very relevant for the evolution as it controls the slope of $\epsilon_2(x)$. We can see this clearly in Fig.~\ref{fig3232g4}, in which the final mass appears to be attracted to the initial mass, approaching it until the asymptotic value of Eq.~\eqref{asympmass} is reached. The results plotted in Fig.~\ref{fig3232g4} also reveal that this behavior towards the initial mass is not symmetric about that value, i.e. the rate of change of the mass is larger for larger $g_4^R$. Hence, for larger asymptotic masses, the convergence to the final value is much faster than for the results below the initial mass. Furthermore, we notice some similarities between this transition and the others, as one sees a faster evolution for smaller $\epsilon_1$ and for smaller $\mu_R^2$. However, the effect is slightly different, since a smaller $\epsilon_1$ essentially contributes to a fast evolution through the terms $\propto 1/\epsilon_1$, but a smaller $\mu_R^2$ removes part of the constant contribution to the mass. This affects the rate of change of the mass somewhat differently as well as the convergence towards the asymptotic mass.
\\

From these numerical results, we were able to find new expressions for the effective mass, which are far more reliable than those obtained in the previous section, given the absence of divergences at late times. In all cases, the results converge to the asymptotic mass, given by Eq.~\eqref{asympmass} and evolve differently depending on the parameters of the system before and after the quench. We also estimate the error in our constant-mass approximation below for the cases under study and conclude that, in spite of the large deviations existing for many situations, there are many relevant parameter values for which one can trust the approximation, which concludes the proof of concept we proposed to do.
\\

\noindent\textbf{Critical analysis of the constant mass approximation}
\\

Concerning our constant mass approximation, we analyze its error in terms of the quantity defined in Eq.~\eqref{error}, $e_u$, by calculating it for all the transitions studied here. We do not expect our results to be trustworthy for all of the cases presented, given the fast evolution of the mass in many. However, we also find several situations in which the error estimate is small, thus making our results reliable.

Regarding \textbf{transition 1}, we find the error to be approximately described by $e_u=3g_4^R/2\epsilon_1$ (in \%), such that a few of the results plotted in Figs.~\ref{fig3212e1} and \ref{fig3212g4} have an error of less than $1\%$, while all except the largest have an error smaller than $10\%$. These case studies justify the approach we have adopted from the beginning.

For \textbf{transition 2}, however, we find that most of the results in Fig.~\ref{fig1232mu} have errors larger than $10\%$. For a value of $\mu_R^2/H^2=0.1$, the error is only smaller than $1\%$ when $g_4^R<2.7\times 10^{-3}$. This changes to $g_4^R<3.7\times 10^{-5}$ for $\mu_R^2/H^2=0.01$. This difference is not surprising, given that we had found a more substantial evolution of the mass for smaller values of $\mu_R^2$. This is also why the results with the smallest error in Fig.~\ref{fig1232g4} are those which have a higher value of $\mu_R^2$. The case $\mu_R^2=H^2/2$, for example, has an error of only $e_u=0.16\%$. The general trend is similar to that of \textbf{transition 1}, with smaller errors for larger $\mu_R^2$ and smaller $g_4^R$.

In the case of \textbf{transition 3}, we report similar error estimates as for the other transitions, again consistent with the error being smaller whenever the evolution is slower. It is possible to find errors smaller than $1\%$ for situations with very small coupling, $g_4^R$, or for large $\epsilon_1$ and $\mu_R^2$. For example, the cases with the rather large $\epsilon_1=0.1$, $\mu_R^2/H^2=0.1$, have errors $e_u<1\%$ if $g_4^R<4.3\times10^{-3}$. An effect that was not present in \textbf{transitions 1} and \textbf{2} takes place here when the difference of masses, or equivalently $\epsilon_2-\epsilon_1$, turns out to be small. In those cases there is a sharp decrease of the error, since the time-dependent terms are suppressed. For example, for $\epsilon_1=0.01$, $\mu_R^2/H^2=0.01$, one finds the error to be $e_u\approx 1\%$ for $g_4^R=1.6\times10^{-2}$, while it is larger than $10\%$ for $g_4^R=10^{-3}$. Other similar examples exist, including situations in which $g_4^R$ is non-perturbative, i.e. of order 1. This is not entirely surprising, given that when $\epsilon_2-\epsilon_1$ is very small, the quench is nearly non-existent.

Furthermore, there is an important point that must be made with respect to the reliability of our approximation. Given that the parameter values for which the error is small are those for which the evolution is suppressed, one could wonder whether our results for the time dependence are accurate at all, i.e. whether they are an improvement to simply saying that, after the quench, one has a constant mass equal to the asymptotic mass. To answer this question, we compute the error with two versions of $u_0(\tau)$. We use the same expression in both cases, but in one we keep the value of the mass constant, while for the other version we substitute for the first approximation of the time dependent mass, $m_1(\tau)$. In all cases studied here, the error is smaller for the second version, indicating that our approximation is converging towards the real evolution of the mass, which is essential for the reliability of the method. Thus, we confirm that we are indeed finding a first approximation to the evolution of the mass and not just its asymptotic value.

\subsection{Negative $m^2$ and symmetry breaking}
\label{sec:Negative}

In this section we study whether non-positive values for $m^2$ are possible and what is the consequence for the spontaneous breaking of the $O(N)$ symmetry of the system.

We begin by re-stating the fact that, in a de Sitter invariant state, IR effects force the effective mass squared, $m^2$, to be strictly positive. This occurs regardless of the sign of $\mu_R^2$, since there always exist solutions to the mass equation for which $m^2$ is positive. This implies that the $O(N)$ symmetry of the system cannot be spontaneously broken, i.e. the only minimum of the effective potential is at $\phi^a=0$.

\begin{figure}[h]
    \centering
    \includegraphics[scale=0.8]{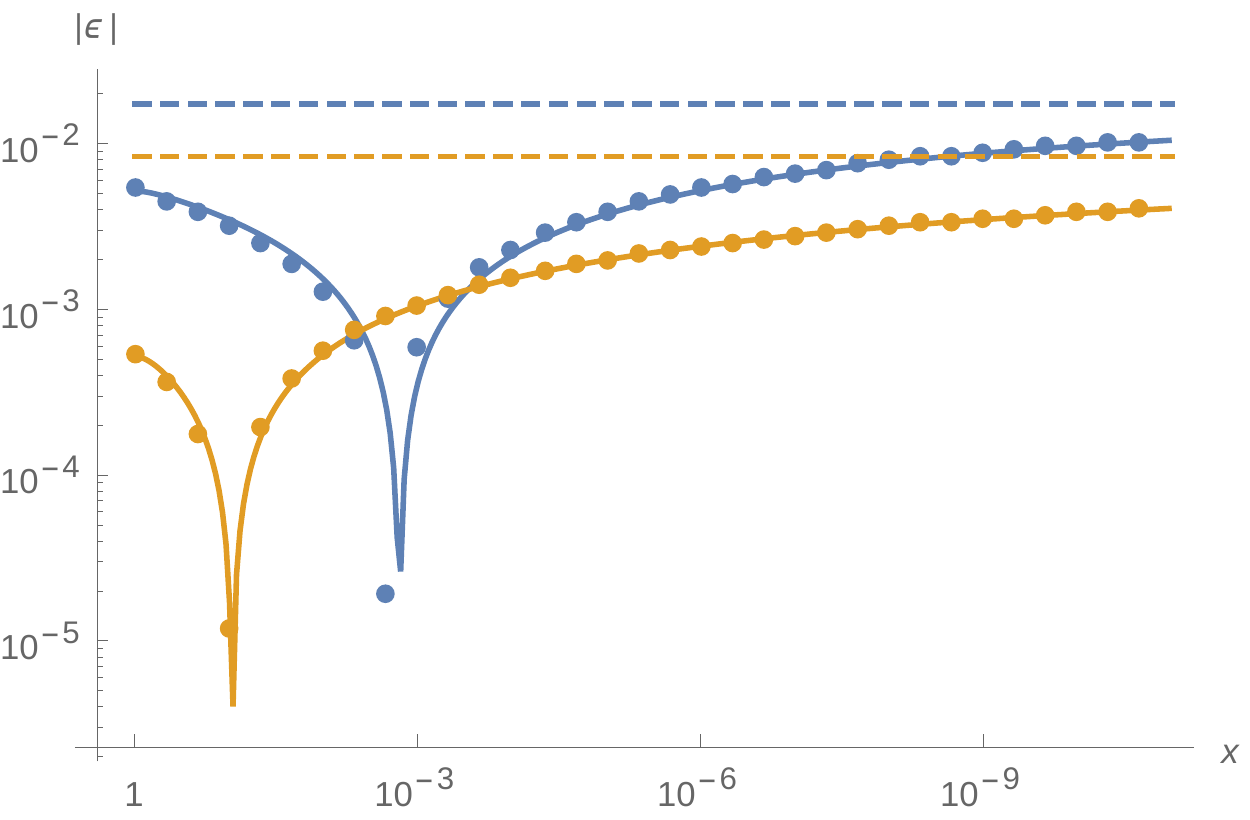}
    \caption{Numerical evolution of $|\epsilon_2(x)|$ for negative $\mu_R^2$ (dotted) as compared to the analytical results (solid) and showing the asymptotic mass (dashed).  Initially $\epsilon_2(x)$ is negative, but evolves towards positive values after some time.}
    \label{fig_negm2}
\end{figure}

For the case of a quench, the scalars are no longer in a de Sitter invariant state, and thus their mass squared may not be strictly positive. While it is true that, asymptotically, the mass squared always converges to the positive value given by the solution of Eq.~\eqref{asympmass}, there is a possibility that it is not always positive throughout the evolution. An analysis of Eq.~\eqref{q3232ren}, for example, reveals that, for values of $\mu_R^2$ that are sufficiently negative, one cannot find solutions for the effective mass squared which are positive. These solutions have been verified with numerical integration and are found to match the analytical results for a negative $m^2$, as shown in Fig.~\ref{fig_negm2}. The absence of IR divergences is due to the quench, as the IR part of the integrals of the power spectrum is dominated by the mass before the quench, $\mu_0^2$, which is positive. The influence of the state before the quench is gradually washed out and thus the mass squared is forced once again to become positive. Therefore, these results indicate that $m^2$ can be negative over the course of the evolution, but only temporarily.

Should the mass squared be negative, however, one expects the $O(N)$ symmetry to be broken and thus that the minimum of the effective potential to change to a non-zero value, i.e. one would have, $\bar\phi^N=H^2\tau^2\bar\sigma\neq0$. The discussion above neglected this factor, which has to be taken into account in the mass equation, Eq.~\eqref{selfconsdS}. We re-write it below in terms of $\phi$ instead of $\bar\sigma$ and $\chi$,
\begin{equation}
m^2=\mu^2+g_4\left[\bar\phi^N(x)^2+i G_\phi(x,x)\right]\,.
\end{equation}
This equation, will, in general, have a different solution due the extra contribution of the term $g_4 \bar\phi^N(x)^2$. Such a contribution is, however, not expected to be present immediately as the quench happens, at $\tau=\tau_0$, as the continuity of the fields imposes $\bar\phi^N=0$ at that time. Thus, the solution to the gap equation at $\tau_0$ remains the same as the one we obtained above, with the extra effect of the background field increasing in time as it evolves towards the minimum of the effective potential. This evolution is difficult to predict within our framework, but it seems clear that the effective mass will approach $m^2(\tau)=0$, as the term $g_4 \bar\phi^2$ cancels the negative $\mu^2$. However, the mass is not expected to remain at this value. If it did, then both the background field, $\bar\phi^N$, and the two-point function $G_\phi(x,x)$ would have to be constant, a situation which only happens in a de Sitter invariant state. But, one already knows from previous arguments that, in such a state, the mass squared must be strictly positive, which it would not be. Therefore, the mass should keep evolving, becoming positive again and eventually reaching the asymptotic value given by Eq.~\eqref{asympmass}, since, in that late-time limit, the background field $\bar\phi^N$ will once again have stabilized at $\bar\phi^N=0$. These arguments are somewhat in disagreement with the results of Ref.~\cite{Boyanovsky:1996rw}, which states that the system should be massless in the late-time limit. Nevertheless, should the mass be zero, it is not clear how one would avoid the IR divergences.

Given the arguments above, we conclude there is the possibility of a transient period in which the mass squared is negative, the duration of which should be calculable from a full numerical evolution of the entire system. We leave that for future work. During this period, the $O(N)$ symmetry of the system is broken to $O(N-1)$, but it is subsequently restored.

\section{Discussion and conclusions}
\label{sec:Discussion}

In this work, we have studied a quantum quench of an $O(N)$ scalar field theory in the background of a de Sitter spacetime. We have obtained the approximate evolution of the effective mass, in the regime in which it is slowly varying. In particular we have derived an expression for the mass in the late-time limit, Eq.~\eqref{asympmass}, which is an accurate limit for the effective mass, even in the general situation not covered by the present approximation. We reproduce that here:
\begin{equation}
m^2_\infty=\mu^2_R+\frac{g_4^RH^2}{16 \pi^2} \left(\frac{m^2_\infty}{H^2}-2\right)\left(\log 4 -1-\Psi\left(\nu_2^\infty-1/2\right)-\Psi\left(-\nu_2^\infty-1/2\right)\right)\nonumber\,,
\end{equation}
with $\nu^\infty_2=\sqrt{9/4-m_\infty^2/H^2}$. Analyzing that limit, we notice that it is independent of the initial mass prior to the quench, in contrast to a similar result in flat spacetime \cite{Hung:2012zr}.

Furthermore, we have obtained analytical expressions for the evolution of the mass, which we summarize in table \ref{tab1} for \textbf{transitions 1}, \textbf{2} and \textbf{3}.

\begin{table}[htpb]

    \heavyrulewidth=.08em
    \lightrulewidth=.05em
    \cmidrulewidth=.03em
    \belowrulesep=.65ex
    \belowbottomsep=0pt
    \aboverulesep=.4ex
    \abovetopsep=0pt
    \cmidrulesep=\doublerulesep
    \cmidrulekern=.5em
    \defaultaddspace=.5em
    \renewcommand{\arraystretch}{1.6}
    \begin{center}
        \small
        \begin{tabular}{Q|q}

            \toprule
            \textrm{Effective mass equation}
            &
            \multicolumn{1}{c}{ Transition}
            \\
            \cmidrule(l){1-2}
            \rowcolor[gray]{0.9}
     \displaystyle m^2=\mu_R^2+\frac{g_4^R H^2}{8\pi^2}x^2\left[\left(\frac{1}{\epsilon_1}-3-2 \log(1-x)\right)(x-2)^2-1\right] &
            \displaystyle \frac{\mu_0}{H}\ll 1\rightarrow \frac{m}{H}\approx\sqrt2
            \\[2mm]

            \cmidrule{1-2}
            \rowcolor[gray]{1.0}

            \displaystyle m^2=\mu_R^2+\frac{g_4^RH^2}{4\pi^2}\left(\gamma_{\rm E}-\frac54+\log 2 +x+\frac{x^4}{4}+\log(1-x)+\frac{1-x^{2\epsilon_2}}{2\epsilon_2} e^{-\frac{3\epsilon_2}{2}}\right)  &
            \displaystyle \frac{\mu_0}{H}=\sqrt2\rightarrow \frac{m}{H}\ll1
            \\[2mm]

            \cmidrule{1-2}
            \rowcolor[gray]{0.9}
            \displaystyle m^2=\mu_R^2+\frac{g_4^RH^2}{8\pi^2}\left(2\gamma_{\rm E}-4+2\log 2+\frac{1}{\epsilon_2}+\frac{\epsilon_2-\epsilon_1}{\epsilon_1\epsilon_2}x^{2\epsilon_2}e^{\frac{2\epsilon_2}{3}(1-x^3)}\right) & 
            \displaystyle \frac{\mu_0}{H}\ll 1\rightarrow \frac{m}{H}\ll1
            \\
             \bottomrule
   
        \end{tabular}
    \end{center}
    \caption{Summary of the solutions to the self-consistent mass in different transitions.}\label{tab1}
    \end{table}

In the table, $x=\tau/\tau_0$ is the ratio between the current value of conformal time, $\tau$, and the initial value, $\tau_0$, at which the quench happened. The parameters $\epsilon_1$ and $\epsilon_2$ are proportional to the initial and final masses and are given by $\epsilon_1\approx \mu_0^2/3H^2$ and $\epsilon_2\approx m^2/3H^2$, respectively.
In all cases, we report an evolution of the effective mass in the direction of the value of the mass before the quench, until it approaches a strictly positive asymptotic value. We confirm this result within our constant mass approximation for many values of the parameters of the system, by showing that the error in the approximation is small. In all other situations, in which the evolution is too fast, we can only be certain about the direction of the initial evolution of the mass and its final value, as per the assumptions of our calculations.

We have also evaluated the possibility of a transition to a negative mass squared and consequent symmetry breaking. We have argued that, should the parameters of the system be such that spontaneous symmetry breaking happens, this stage will be
transient, with the symmetry being restored after a certain time. Within our approximations, that time interval cannot be calculated and hence its evaluation is left for future work.

\para{Implications for cosmology}One of our original motivations was the direct application of the quench to fast transitions during inflation. If one interprets the scalars under study here as the perturbations 
of the inflaton field, the effect of the quench can be seen by calculating the power spectrum from Eq.~\eqref{bogtrans}.

\begin{figure}[h]
    \centering
    \includegraphics[scale=0.9]{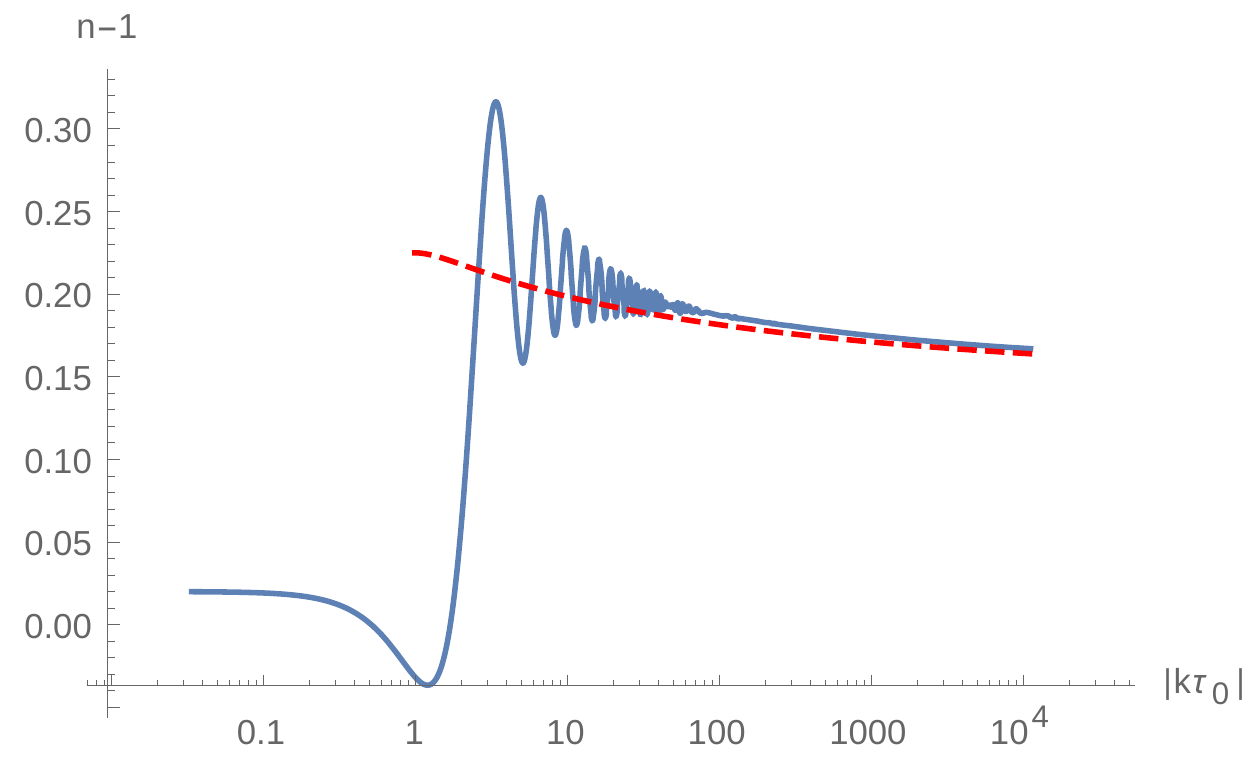}
    \caption{The spectral index (solid) 
    as a function of scale $k$ for the transition with parameters $\epsilon_1=0.01$, $g_4^R=0.1$, $\mu^2_R=0.2$, shown at a time in which all scales have become super-horizon. Also shown is the value of $2\epsilon_2$ (dashed), representing the effective mass via $\epsilon_2\approx m^2/3H^2$ and evaluated at the time each scale exited the horizon, $\tau=k^{-1}$.}
    \label{figns}
\end{figure}

Another key quantity is the spectral index, which can be derived from the power spectrum, $\mathcal{P}=k^3\langle\phi^2\rangle$, via
\begin{equation}
    n-1=\frac{{\rm d}\log \mathcal{P}}{{\rm d}\log k} \ .
\end{equation}
Evaluating the spectral index at the end of inflation, one would see an abrupt change in its value, occurring approximately at the scale $k_0\sim\tau_0^{-1}$, accompanied by small oscillations for $k>k_0$, as can be seen in Fig.~\ref{figns}. This is because the spectral index depends on the mass of the field at the time a certain scale left the horizon, and therefore will be sensitive to when the quantum quench occurs.

This situation is quite similar to what is described in Ref.~\cite{Joy:2007na}. However, given that our results do not take slow-roll into account, nor do we attribute the accelerated expansion to the effects of our scalars, the tendencies described here may not be realized in practice.

In any case, we have shown in this paper that it is possible to solve for the dynamics of a scalar field theory after a quantum quench in de Sitter spacetime, which is a very important first step towards the application to inflation. Beyond what we have done here, a full numerical evolution of the mode equation, Eq.~\eqref{EOMWF}, would be required, as well as the solution of the background equations, as those are also affected by the quench.

Another interesting application would be to the study the effect of spectator fields in inflation. It would be particularly interesting to study the quench to a negative mass, described in section \ref{sec:Negative}, to check if that period can last for long enough to destabilize the slow-roll expansion and potentially end inflation.\\

\para{Summary}We have introduced a new method to study fast transitions in de Sitter spacetime using the large-$N$ technique. We 
have obtained an approximate solution to the dynamics of the system, which we believe to include most of the relevant features of the full solution, including the time dependence of the mass and its asymptotic value. We have also pointed to future directions, including a more direct application to inflation using numerical methods.

\acknowledgments{We would like to thank Tim Clifton, Sophia Goldberg, Karim Malik, David Mulryne, Viraj Sanghai and Tommi Tenkanen for useful discussions and comments on a draft version of the paper. We would especially like to thank Ling-Yan Hung for many discussions and ideas on the subject of quantum quenches and for help in the initial stages of this project. PC acknowledges support from a Queen Mary Principal's Research studentship and a Bolsa de Excel\^encia Acad\'{e}mica of the Funda\c{c}\~{a}o Eug\'{e}nio de Almeida}. RHR acknowledges support from the Science and Technology Facilities Council grant ST/J001546/1 and the hospitality of the Perimeter Institute for Theoretical Physics where this work was initiated.

\bibliographystyle{JHEPmodplain}
\bibliography{references}

\end{document}